\documentclass[10pt,aps,prd,twocolumn,showpacs,groupedaddress,nofootinbib]{revtex4-1}
\usepackage[english]{babel}
\usepackage{newtxtext,newtxmath}
\usepackage{graphicx}
\usepackage{amsmath}
\usepackage{upgreek}
\usepackage{units}
\usepackage{booktabs}
\usepackage{xspace}
\usepackage{xcolor}
\definecolor{darkred}{rgb}{0.3,0,0}
\definecolor{darkblue}{rgb}{0,0,0.3}
\definecolor{firebrick}{rgb}{0.5,0.125,0.125}
\definecolor{darkgreen}{rgb}{0,0.3,0}
\usepackage[colorlinks=true,linkcolor=firebrick,citecolor=darkgreen,urlcolor=darkblue]{hyperref}
\usepackage[capitalize]{cleveref}

\graphicspath{{figs/}}

\newcommand{\funk}{\mbox{\textsc{Funk}}\xspace}

\begin{document}

\title{Limits from the \textsc{Funk} Experiment on the Mixing Strength of Hidden-Photon Dark Matter in the Visible and Near-Ultraviolet Wavelength Range}

\def\affkit{Institute for Nuclear Physics, Karlsruhe Institute of Technology \includegraphics[height=1.55ex]{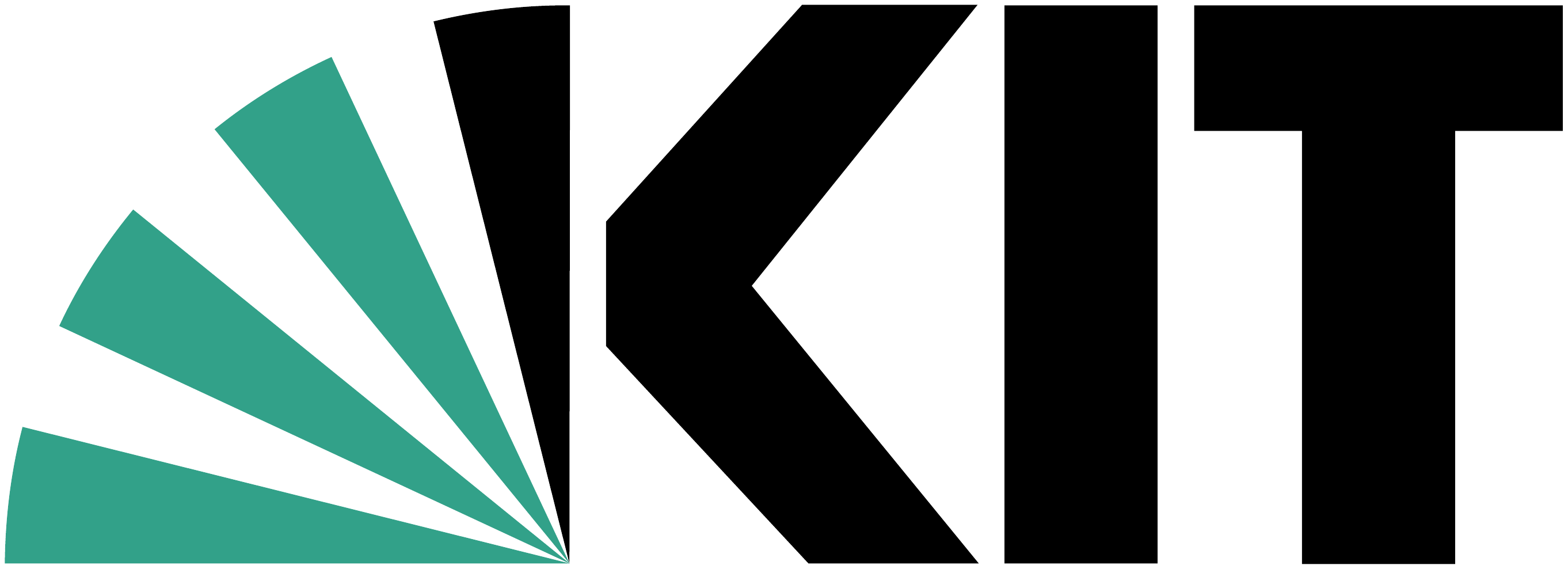}, Germany}
\def\affcern{Physics Department, CERN, Geneva, Switzerland}
\def\affitp{Institute for Theoretical Physics, Heidelberg University, Germany}
\def\affdesy{Deutsches Elektronen Synchrotron (DESY), Hamburg, Germany}
\def\affhum{Department of Physics, Humboldt University, Berlin, Germany}
\def\affzar{Department of Theoretical Physics, University of Zaragoza, Spain}

\author{Arnaud Andrianavalomahefa}
\affiliation{\affkit}

\author{Christoph M.\ Sch\"afer}
\affiliation{\affkit}

\author{Darko Veberi\v{c}}
\email[]{darko.veberic@kit.edu}
\affiliation{\affkit}

\author{Ralph Engel}
\affiliation{\affkit}

\author{Thomas Schwetz}
\affiliation{\affkit}

\author{Hermann-Josef~Mathes}
\affiliation{\affkit}

\author{Kai~Daumiller}
\affiliation{\affkit}

\author{Markus~Roth}
\affiliation{\affkit}

\author{David Schmidt}
\affiliation{\affkit}

\author{Ralf Ulrich}
\affiliation{\affkit}

\author{Babette~D\"obrich}
\affiliation{\affcern}

\author{Joerg~Jaeckel}
\affiliation{\affitp}

\author{Marek~Kowalski}
\affiliation{\affdesy}
\affiliation{\affhum}

\author{Axel~Lindner}
\affiliation{\affdesy}

\author{Javier~Redondo}
\affiliation{\affzar}

\author{\includegraphics[width=20mm]{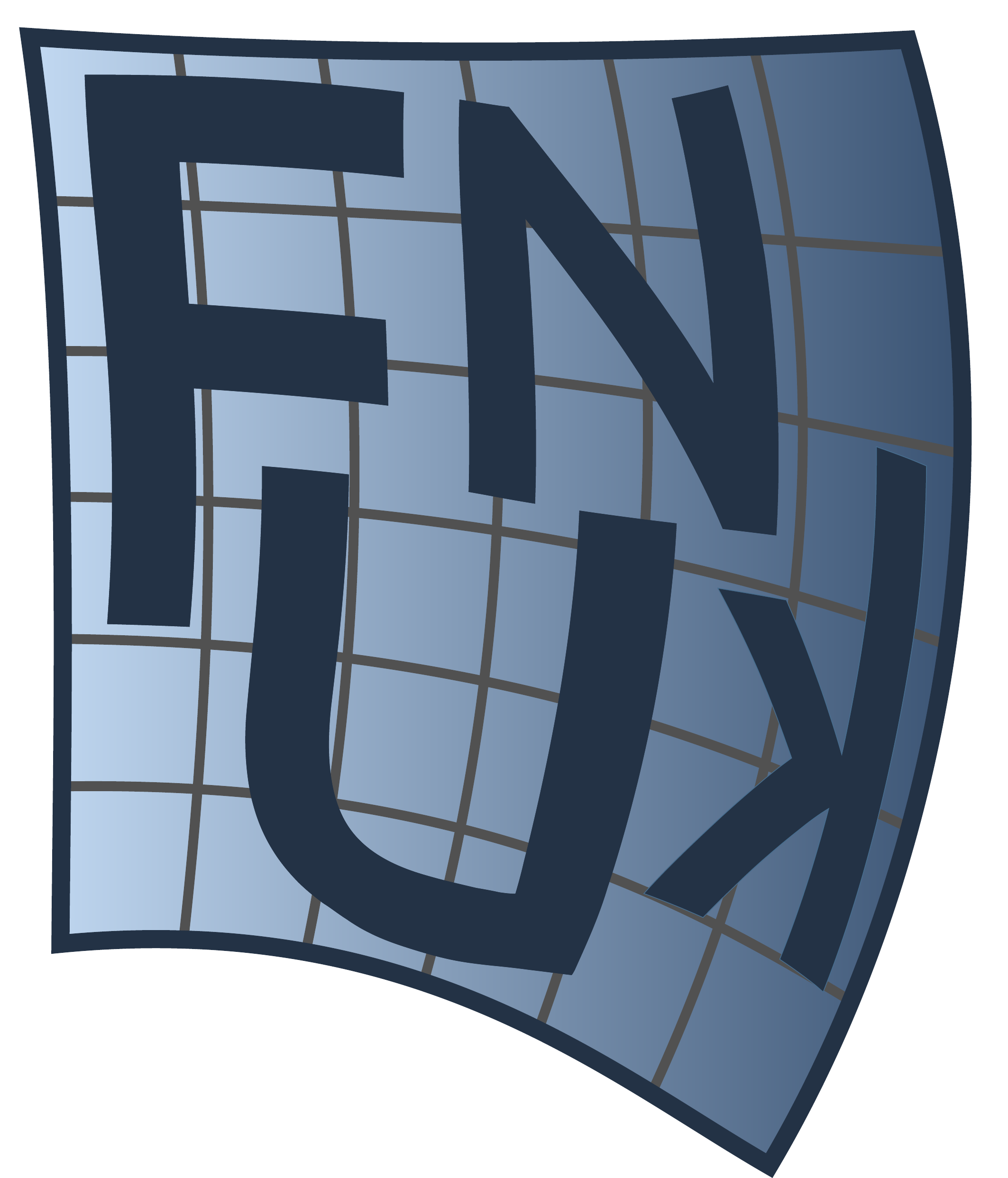}}
\collaboration{The \funk Experiment}

\begin{abstract}
We present results from the \funk experiment in the search for hidden-photon dark matter.
Near the surface of a mirror, hidden photons may be converted into ordinary photons.
These photons are emitted perpendicularly to the surface and have an energy equal to the mass of the dark matter hidden photon.
Our experimental setup consists of a large, spherical mirror with an area of more than 14\,m$^2$, which concentrates the emitted photons into its central point.
Using a detector sensitive to visible and near-UV photons, we can exclude a kinetic-mixing coupling of stronger than $\chi\approx10^{-12}$ in the mass range of 2.5 to 7\,eV, assuming hidden photons comprise all of the dark matter.
The experimental setup and analysis used to obtain this limit are discussed in detail.
\end{abstract}

\pacs{dark matter, hidden photons, spherical mirror, PMT, dark current}

\maketitle

\section{Introduction}

Constituting approximately 85\% of its matter and 25\% of its total energy content~\cite{Akrami:2018vks}, dark matter (DM) is possibly one of the central puzzles in our quest to understand the Universe.
Structure formation indicates that DM behaves, to a very good approximation, like a cold and pressureless gas, which suggests that it is made from slow, non-relativistic particulates.
The mass of these particulates is, however, essentially unknown.
They could be elementary particles with masses as low as ${\sim}10^{-22}$\,eV, but they could also be macroscopic objects, such as black holes, with masses as large as ${\sim}10\,M_\text{sun}$.

One of the most promising ways to answer this dilemma is a direct detection of the DM particles in a laboratory environment.
In this paper, we present the results of exactly such an experiment.

In particular, our experiment, ``Finding U(1)'s of a Novel Kind'' (\funk), searches for ``dark'' or ``hidden'' photons (HPs) in a light range of masses around ${\sim}$eV.

Following Occam's razor, this option is especially appealing since it extends the Standard Model (SM) gauge group by only a single U(1) gauge factor giving rise to the HP.
Taking all SM particles to be uncharged under the new U(1) makes the extra gauge boson ``hidden''.
The only interaction then arises from a coupling to the ordinary photon through kinetic mixing~\cite{Holdom:1985ag}.
Analogous to neutrino oscillations, this mixing leads to photon-HP oscillations.
For the oscillations to be observable, the HP needs to have a non-vanishing mass that can arise either via the St\"uckelberg or the Higgs mechanism (see~\cite{Jaeckel:2013ija} for a recent review).
The relevant Lagrangian is given by
\begin{align}
\mathcal{L} &\supset
  - \frac{1}{4} F_{\mu\nu} F^{\mu\nu}
  + J^\mu A_\mu
\nonumber
\\
  &~~~~
  - \frac{1}{4} X_{\mu\nu} X^{\mu\nu}
  - \frac{\chi}{2} F_{\mu\nu} X^{\mu\nu}
  + \frac{m_{\gamma'}^2}{2} X_\mu X^\mu.
\end{align}
As usual, $F_{\mu\nu}$ is the field strength of ordinary photons ($A_\mu$), and $X_{\mu\nu}$ is the analogue for hidden photons ($X_\mu$).
The two new parameters of the model are the kinetic mixing $\chi$ and the HP mass $m_{\gamma'}$.
Explicit models often predict small or even tiny values for the coupling $\chi$~\cite{Holdom:1985ag,Dienes:1996zr} (see~\cite{Jaeckel:2013ija} for more references), which is very much in line with this particle being ``hidden''.

Over the past few years, this simple model has received considerable interest and inspired a number of searches (see, e.g.~\cite{Hewett:2012ns,Essig:2013lka,Battaglieri:2017aum,Alemany:2019vsk,Beacham:2019nyx}). 

In particular, two mass ranges have been focused on.
MeV to GeV mass-scale photons could act as messengers of DM~\cite{ArkaniHamed:2008qn} but have also been proposed to address the discrepancy between the anomalous magnetic moment of the muon ($g{-}2$) and its value as predicted by the SM ~\cite{Bennett:2006fi,Pospelov:2008zw}.

Another possibility explored in many laboratory-scale experiments, is that the HP is very light, typically at or below the eV scale.
Such HPs are intriguing not only because of their simplicity but also because they have been shown to be a viable DM candidate~\cite{Nelson:2011sf,Arias:2012az,Graham:2015rva,Agrawal:2018vin,Dror:2018pdh,Co:2018lka,Bastero-Gil:2018uel,Long:2019lwl}.
This is the case we are interested in.

As already mentioned, if such HPs were to exist, they would be able to oscillate into ordinary photons.
Thus, even without assuming that HPs are DM, searches can be performed for HPs using light-shining-through-walls experiments~\cite{Woollett:2015gma,Betz:2014wie,Ehret:2010mh} or observations of the sun~\cite{Redondo:2015iea,Schwarz:2015lqa}.

To detect HPs in the MHz and GHz range, ``classical'' cavity searches~\cite{Sikivie:1983ip} may be performed~\cite{Nguyen:2019xuh}, or a reinterpretation of cavity searches can be made when the cavity is inserted into a magnetic field to search for the axion~\cite{Arias:2012az}.
At even lower frequencies, ``radio'' searches for HP DM have been put forward~\cite{Silva-Feaver:2016qhh,Arias:2014ela,Chaudhuri:2014dla}.

\begin{figure}[t]
\centering
\includegraphics[width=\linewidth]{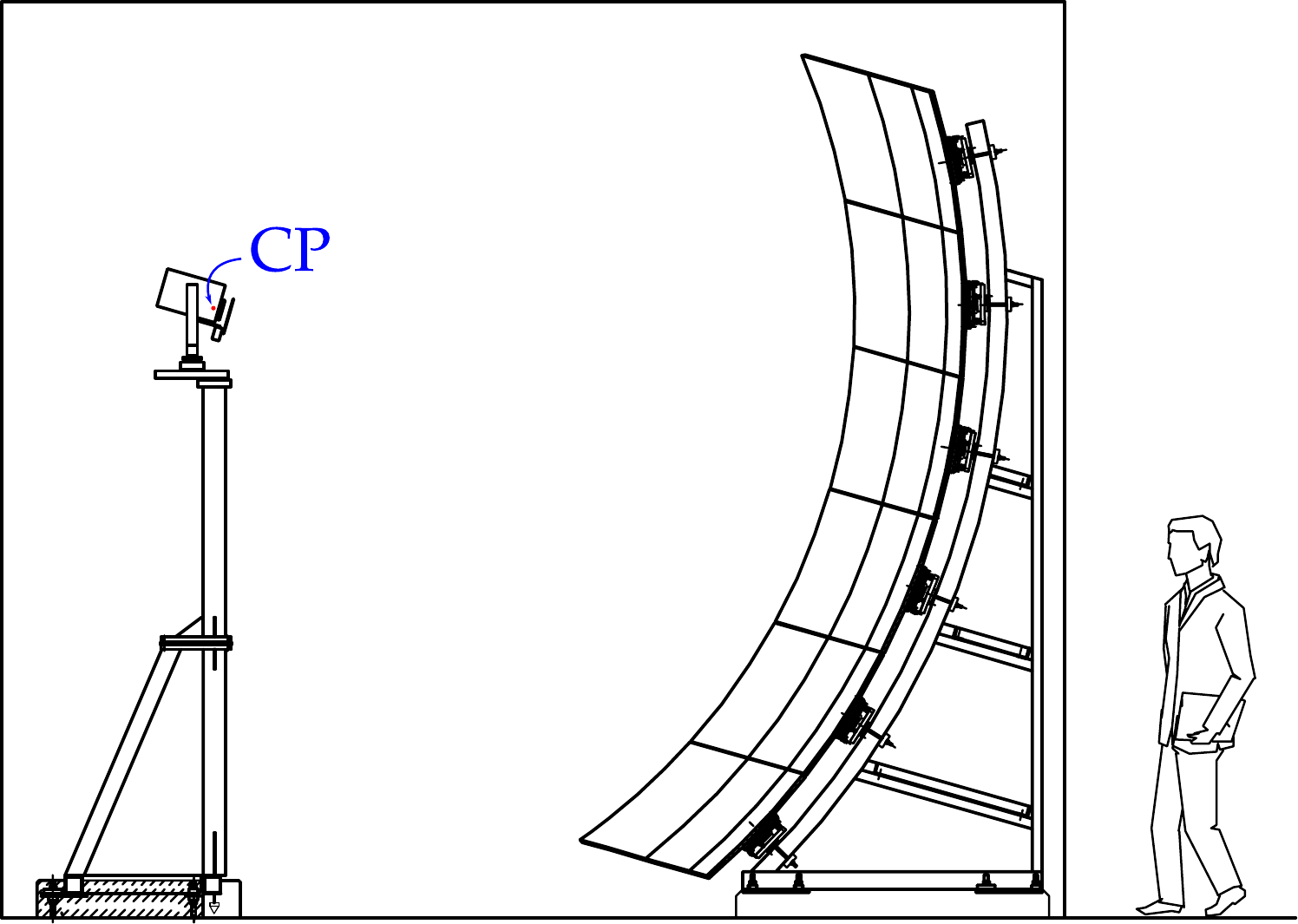}
\caption{Schematic drawing of the \funk experimental setup.
The detector (photomultiplier) is placed on top of a steel pillar (left) so that the sensitive surface is located exactly in the central point (CP) of the mirror (right).
The whole experimental area (indicated with a rectangle) is enclosed with opaque materials.
On the right, a human figure is also shown for scale. (Slightly adapted from~\cite{Experiment:2017icw,Experiment:2017kmm}.)}
\label{f:setup}
\end{figure}

Another elegant method to detect HPs by using large, spherical mirrors, the so-called ``dish antennas,'' has been proposed~\cite{Horns:2012jf,Jaeckel:2013eha}.
In contrast to cavity searches, where each single measurement only probes a mass range equal to the width of the employed cavity, this is, in principle, a broadband method that can be employed to test a wide parameter space when appropriate detectors are employed.
The search with the dish antenna can be understood as follows: Owing to their (tiny) effective coupling to electrons, HPs can induce oscillations of the free electrons in a conductor.
As a consequence of this excitation, ordinary photons can be emitted from the surface of the conductor.
This emission may be increased by choosing a conductor with high reflectivity and forming it into a spherical mirror to facilitate the collection of photons.
The emitted ordinary photons can then be collected at the central point (CP) of the spherical mirror, whereas no such focusing would take place for the background photons with random arrival directions.
This (geometrical) enhancement comes from the fact that the ordinary photons, generated by HP DM, are emitted almost perpendicularly to the surface with only a tiny directional modulation~\cite{Jaeckel:2015kea} due to Earth's relative velocity (or the relative velocity of an experiment) with respect to the rest frame of the HP DM.

The time-averaged power $P$ of ordinary photons emitted by the HPs emanating from the mirror towards the CP is
\begin{equation}
P = \chi^2 \, \langle\cos^2\theta\rangle \, \rho_\text{DM} \, A_\text{mirr},
\label{e:power}
\end{equation}
where $\theta$ is the angle of polarization of the HPs with respect to the mirror plane, and the brackets $\langle\cdot\rangle$ indicate the average over the local population.
Moreover, $\rho_\text{DM}$ is the local DM density, and $A_\text{mirr}$ is the area of the mirror.
Accordingly, a large $A_\text{mirr}$ can greatly enhance the collected power and increase the sensitivity of the setup.

The \funk experiment exploits exactly the aforementioned dish-antenna method.
In the following, we report our results from an ${\sim}1$ month-long data-acquisition campaign, which took place in 2019.
Preliminary results and calibrations were presented in a number of proceedings~\cite{Veberic:2015yua,Dobrich:2015tpa,Experiment:2017icw,Experiment:2017kmm}, some of which we repeat here for convenience and self-containment purposes.

Besides our experiment, three other experiments~\cite{Suzuki:2015sza,Knirck:2018ojz,Brun:2019kak} have published results based on the dish-antenna method.
However, the surface of our mirror is about one order of magnitude larger, which on the one hand strongly enhances the experimental sensitivity while on the other hand increases the complexity of this experiment.
Even though the principles behind the dish-antenna method are quite simple, \funk required a careful design and a number of iterations to arrive at the final sensitivity reported in this paper.

Owing to a comparatively large mirror size and relatively low noise of the chosen detector, we are able to set limits on HP coupling on the order of $\chi<10^{-13}$, which is comparable to the strongest astrophysical constraints existing for HP DM in this mass range.

The paper is structured as follows: \cref{s:funk_setup} describes the experimental setup including mirror alignment, external-light shielding of the experimental area, mechanics, and the data-acquisition and muon-monitoring systems.
\cref{s:analysis} describes the data analysis, including the observation of a ``memory effect'' in the photon sensor, the understanding of which crucially affected the analysis procedure.
In \cref{s:results}, based on the non-observation of a (significant) signal above background, we set upper limits on the mixing parameter $\chi$ for the specific HP mass range corresponding to our detector.

\section{Experimental setup}
\label{s:funk_setup}

\subsection{Mirror}

\begin{figure}[t]
\centering
\def\figh{0.483}
\includegraphics[height=\figh\linewidth]{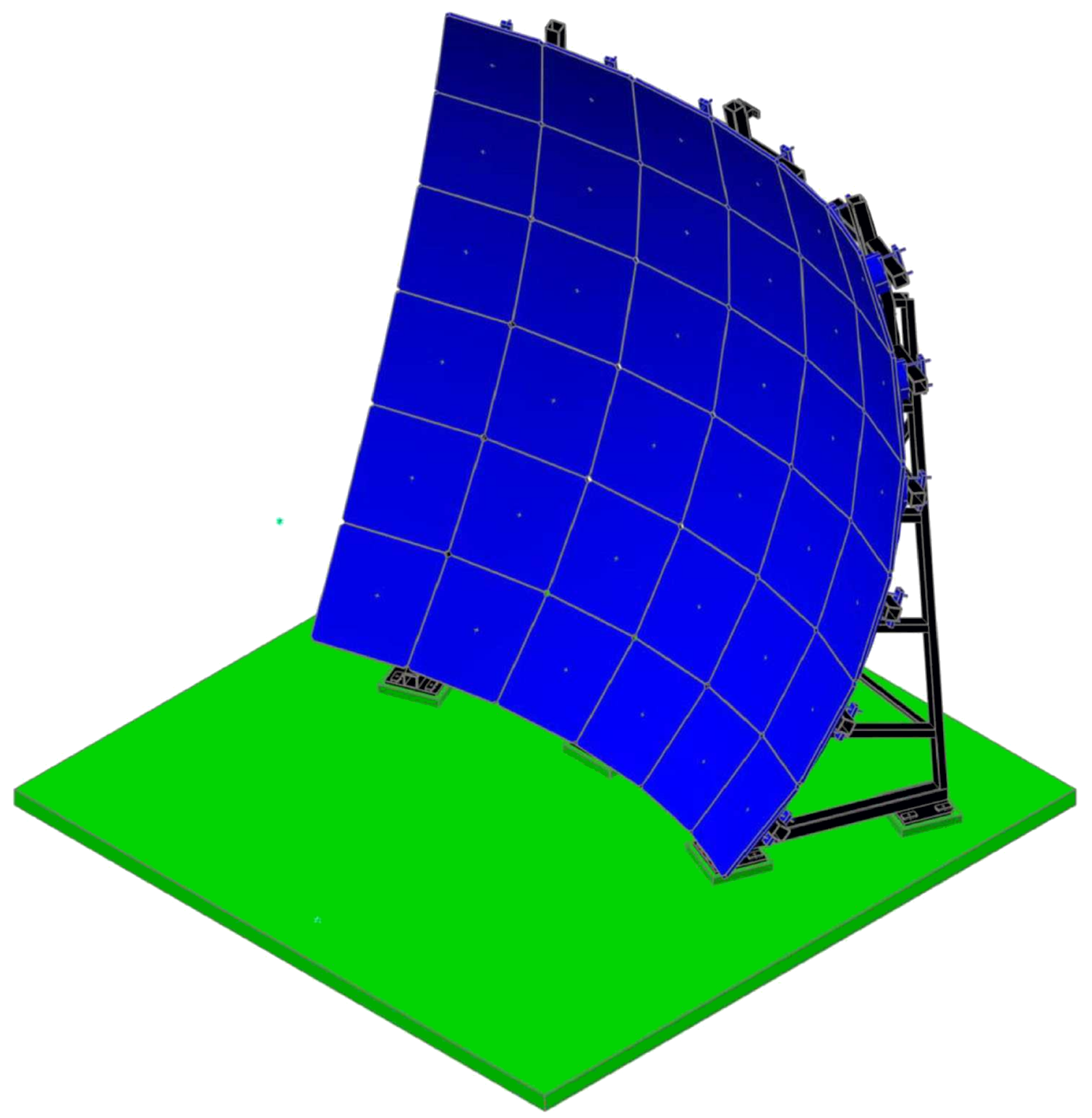}\hfill
\includegraphics[height=\figh\linewidth]{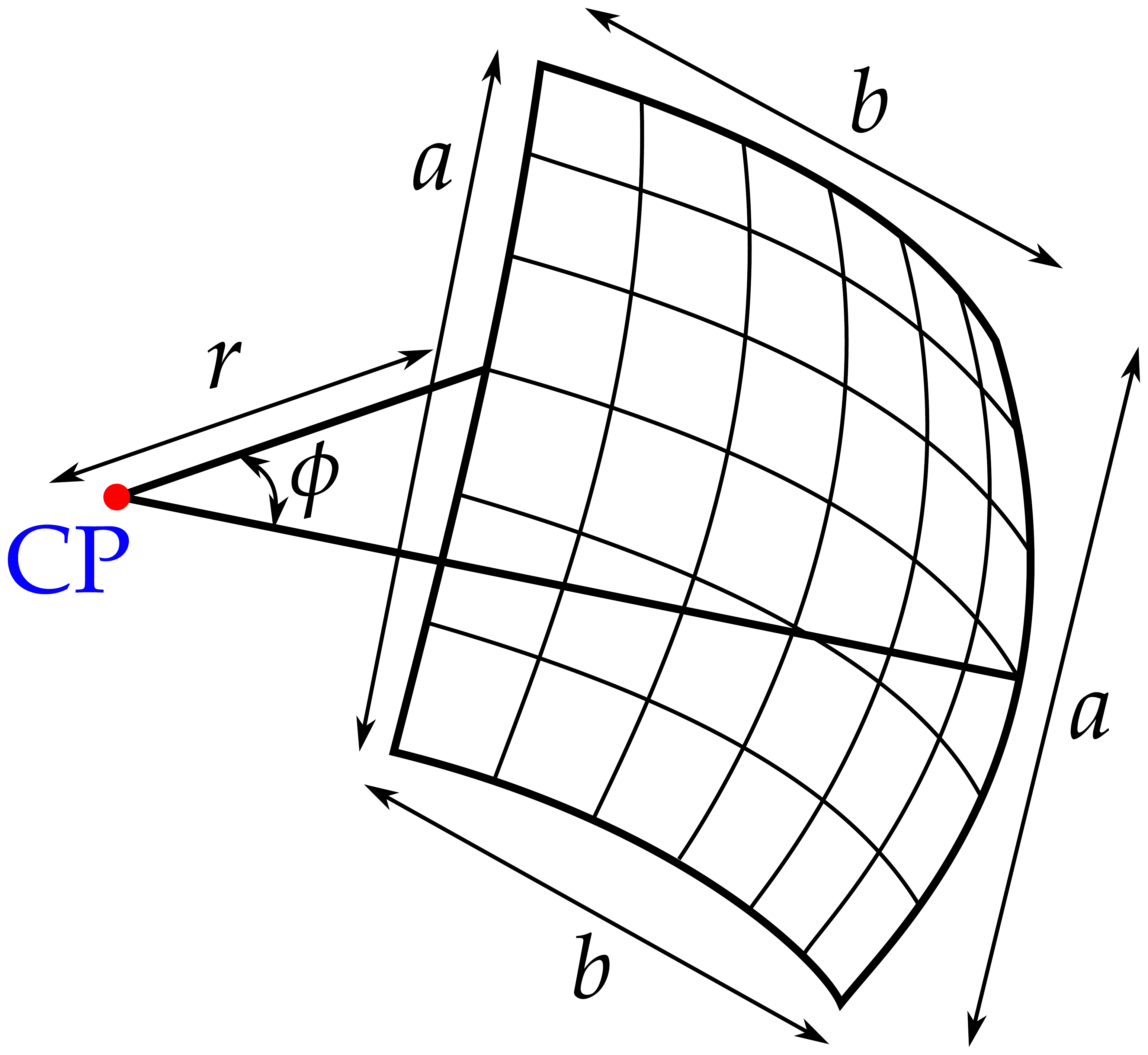}
\caption{\emph{Left:} Placement of the assembled mirror (blue) and its support structure (black) in the experimental area (green).
\emph{Right:} Illustration of the dimensions of the spherical mirror, where $r=3.40$\,m, $a=3.72$\,m, $b=3.70$\,m, and $\phi=66^\circ$, resulting in a total mirror area of $A_\text{mirr}=14.56$\,m$^2$.
Adapted from~\cite{Veberic:2015yua}.}
\label{f:mirror}
\end{figure}

As a suitable surface for the conversion of HPs, we employ a prototype mirror (see \cref{f:setup}) of the Pierre Auger Observatory~\cite{Abraham:2004dt,ThePierreAuger:2015rma}.
As shown in \cref{f:mirror}, the spherical mirror is built from $6\times6$ approximately rectangular pieces, each consisting of an Al base with a thin surface of an Al-Mg-Si-alloy.
The dimensions of the mirror are indicated in \cref{f:mirror}.
For our purposes, the most relevant feature is the large reflecting area $A_\text{mirr}=14.56$\,m$^2$.
See~\cite{Veberic:2015yua} for more details.

\begin{figure}[t]
\centering
\includegraphics[width=\linewidth]{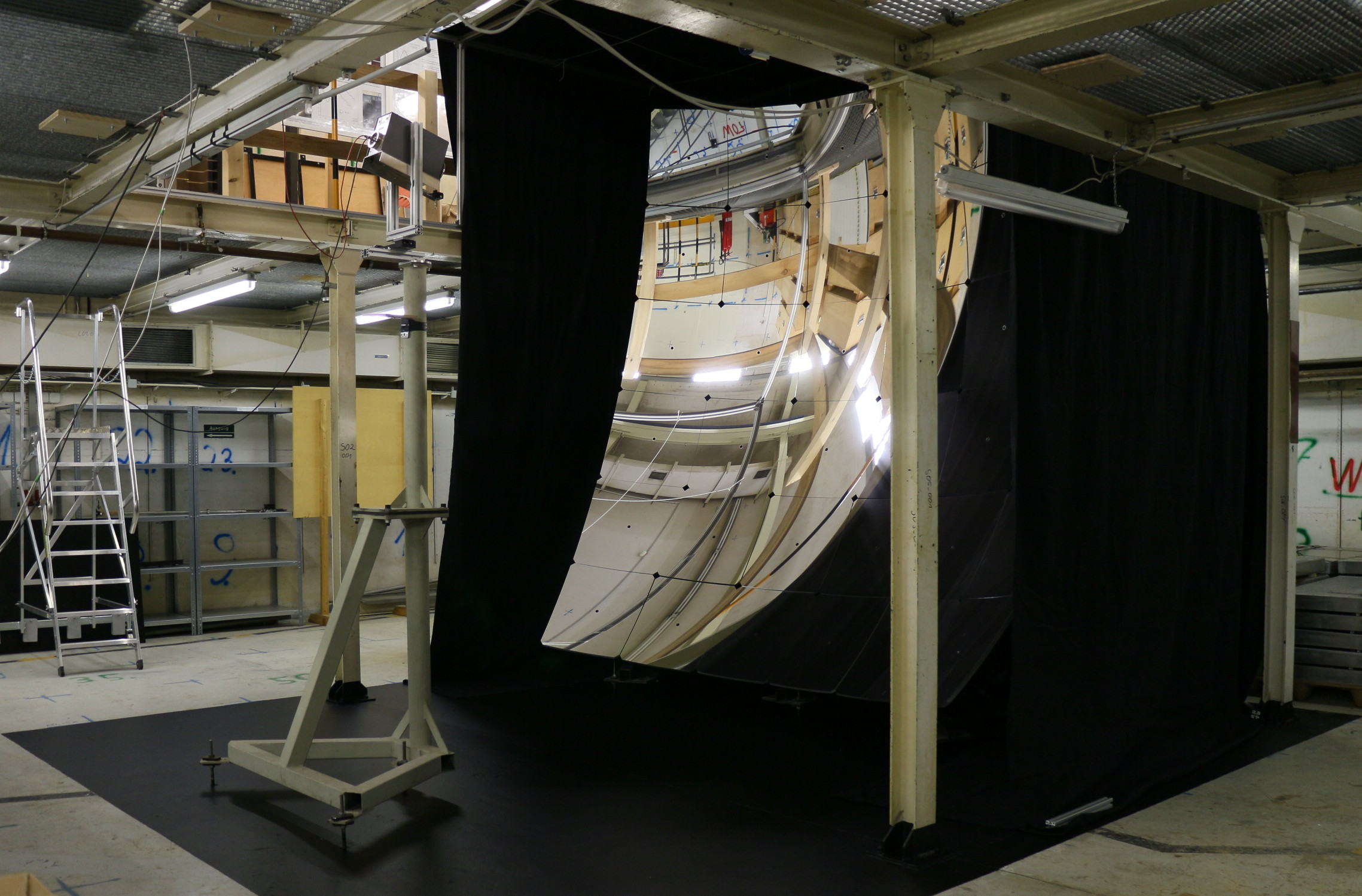}
\caption{Photograph of the experimental area, where the black-painted floor can be seen with the PMT-camera pillar on the left side and the spherical mirror on the right side.
The black-cotton curtain, shown here behind the mirror in a partially installed state, eventually enclosed the whole area and was covered with another layer of black-polyethylene foil.
Taken from \cite{Experiment:2017icw,Experiment:2017kmm}.}
\label{f:area}
\end{figure}

We located the experiment in a windowless, air-conditioned experimental hall with the approximate dimensions $18\,\text{m}\times14\,\text{m}\times9$\,m and with 2\,m thick concrete walls.
The main experimental zone takes up a volume of $4.35{\times}4.98{\times}4.30$\,m$^3$ (see \cref{f:area}) and has been for light-shielding purposes enveloped with a double-layer of black polyethylene sheets with a thickness of $120\,\upmu$m stretched over thick cotton curtains.
Furthermore, we treated the floor within this confined area with a non-reflective and non-glossy black paint.

The assembly of the spherical mirror first followed the standard procedure for construction of the Auger fluorescence telescopes~\cite{Abraham:2009pm}.
Each mirror segment has an adjustable pointing direction and can be radially moved forward/backward by a few centimeters.
To capture as many of the produced photons as possible, the spherical mirror must focus them into the CP with a small spot size\footnote{We stress that the produced photons are concentrated in the CP and not in the usual focal point of the mirror.
This is due to the perpendicular emission from the spherical surface~\cite{Horns:2012jf}.}.
The process to quantify the quality of this focus was already described in~\cite{Veberic:2015yua,Experiment:2017icw}; however, we briefly summarize it again here. 

Any light ray emanating from a source located at the CP of an ideal spherical mirror hits the mirror surface at a right angle.
The ray is therefore by the spherical mirror reflected exactly back into the CP.
On the other hand, a slight lateral offset of the source results in reflected rays focusing at a point mirrored across the CP.
With a suitably small source, the mirror segments can therefore be aligned by focusing their reflected light.
For this purpose, we used a platform consisting of a frosted-glass pane, which served as a suitable projection screen, and a yellow-green LED as the light source.

\begin{figure*}[t]
\centering
\includegraphics[width=0.9\linewidth]{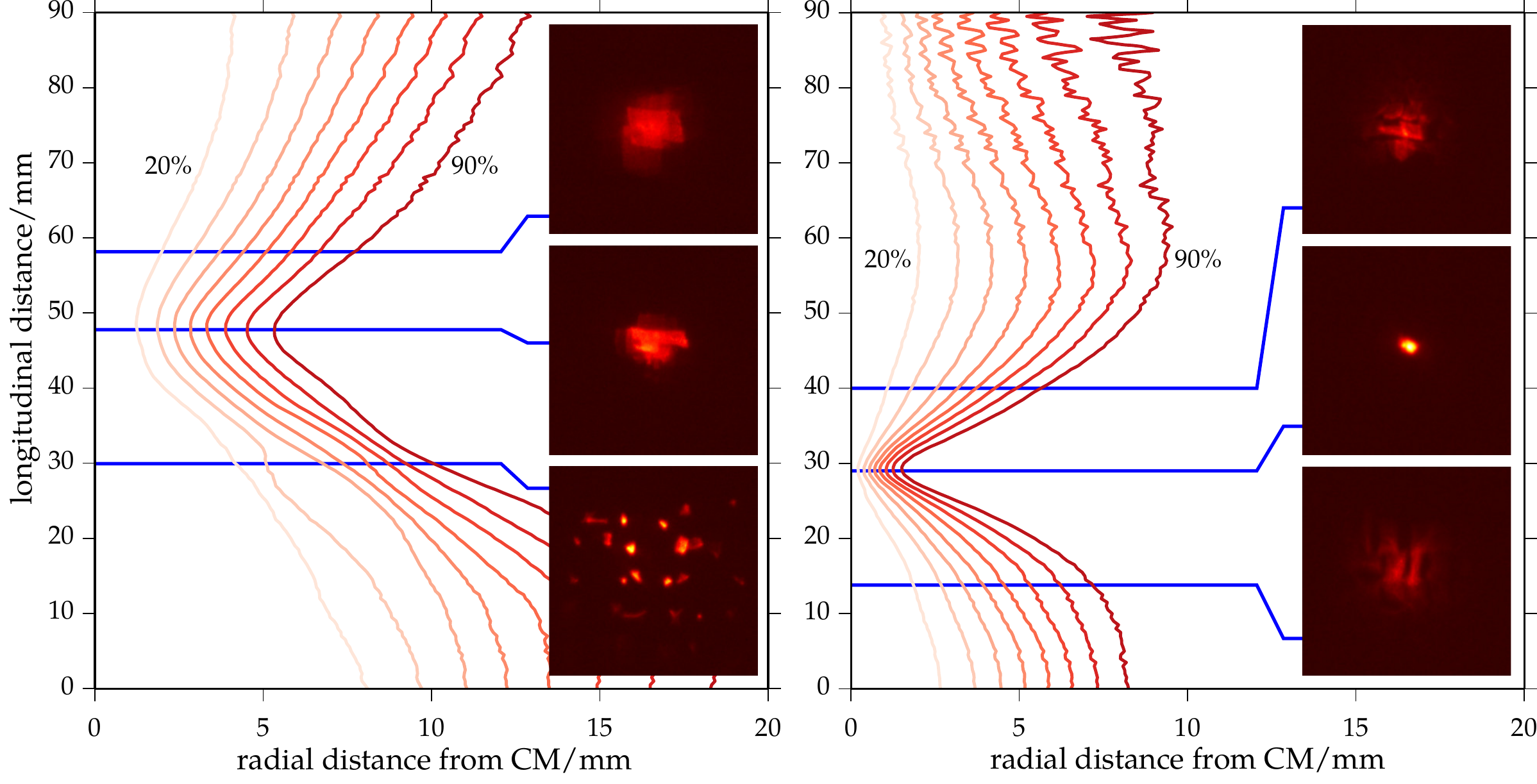}
\caption{Illustration of the point-spread function of the mirror before and after the realignment of the mirror segments:
\emph{Left:} Contours of radial distance (i.e.\ perpendicular to the optical axis of the mirror) from the image center of mass (CM) containing certain fractions of the total light (from 20\% to 90\% in steps of 10\%).
The blue lines mark the longitudinal distance (i.e.\ along the optical axis) of the three images of the spot, shown as insets to the plot.
\emph{Right:} Similar to the plot on the left, but after the dedicated realignment campaign.
The radial size of the spot containing 90\% of the light has shrunk from 6 to 2\,mm (see the two middle insets).
Taken from ~\cite{Veberic:2015yua}, see also~\cite{Experiment:2017icw}.}
\label{f:mirror_align}
\end{figure*}

We employed a standard CCD camera to observe this reflection pattern on the frosted-glass screen.
By moving the screen along the optical axis of the mirror, we then obtained cross-sectional images of the converging light beams from the individual mirror segments.
The position corresponding to the smallest spot size and best sharpness of these images is then taken as the nominal CP of the mirror, where we locate the detector.

In \cref{f:mirror_align}-(left), we present the result of such a scan around the CP of the initial setup.
The insets show the images obtained at different longitudinal positions of the screen, where the central inset corresponds to the position with the smallest spot size.
This is confirmed by the contour lines, which indicate the fraction of the light contained within a circle of a given radius from the image center-of-mass (CM).
The smallest spot for the 90\% quantile is observed at a radius $r\approx6$\,mm.
Potential for improvement can be seen when observing the lower inset of \cref{f:mirror_align}-(left).
Here, individual mirror segments are in good focus but unfortunately with all the foci in different positions.
Therefore, a fine realignment of the pointing and the radial distance of each individual segment was required to achieve the smallest possible spot size (more details can be found in \cite{Veberic:2015yua}).
This procedure enabled us also to compensate for the manufacturing variance in the radius of curvature of individual mirror segments.
The improvement of the spot size (and thus the accuracy of the CP position) is illustrated in \cref{f:mirror_align}, where the size of the spot has been reduced by a factor of 3 (to only 2\,mm for the 90\% quantile).
Note that the spot size is now much smaller than the active area of the detector described in the next section.

\subsection{Detector and calibration}
\label{s:detector}

\begin{figure}[t]
\centering
\includegraphics[width=\linewidth]{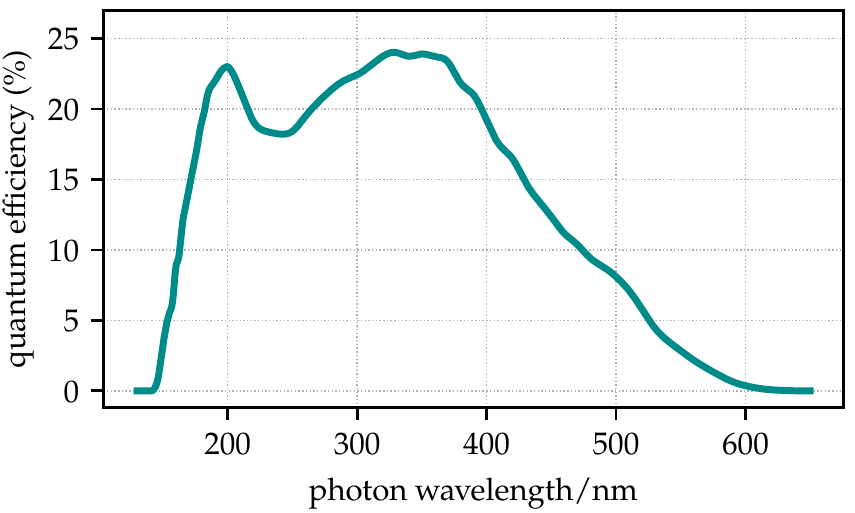}
\caption{Quantum efficiency of our specific ET\,9107QB photomultiplier, as obtained by a dedicated measurement of the manufacturer's laboratory.}
\label{f:q_eff}
\end{figure}

The measurements reported in this paper were made by a low-noise photomultiplier tube 9107QB produced by ET Enterprises~\cite{et-manual}.
This photomultiplier tube (PMT) has a sensitivity in the optical and near-ultraviolet range (wavelengths between 150 and 630\,nm) and reaches peak quantum efficiency at $q_\text{eff}(330\,\text{nm})=24\%$ (see \cref{f:q_eff}).
Additionally, when operating at high gain, it exhibits low levels of noise and good single-photoelectron (SPE) resolution, thus making it an excellent detector for photon-counting applications.

\begin{figure}[t]
\centering
\includegraphics[trim={0 320 0 80},clip,width=\linewidth]{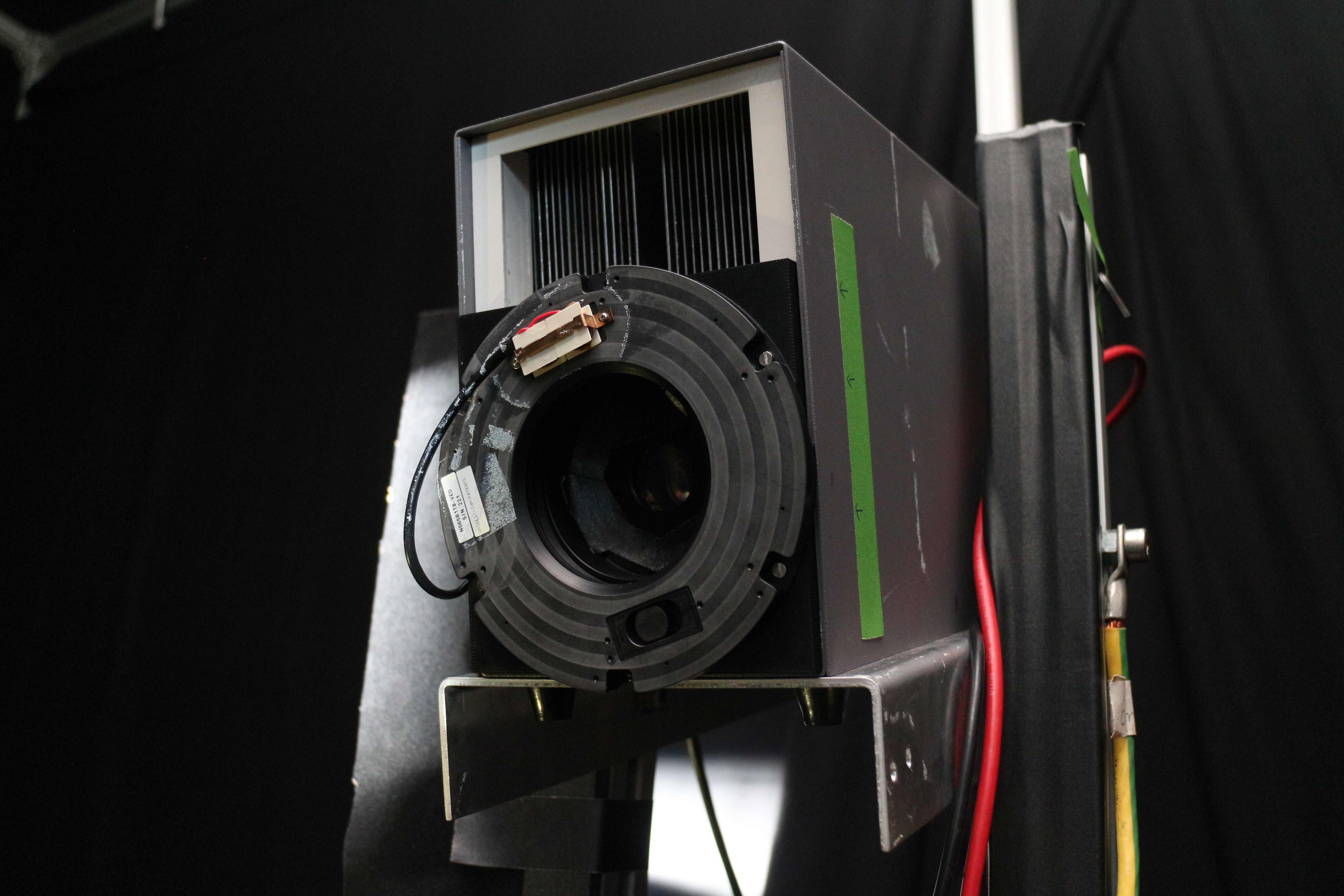}
\caption{Photograph of the camera setup. The PMT is housed in the FACT50 enclosure (gray box).
The photocathode of the PMT can be seen through the iris shutter with an inner diameter of 65\,mm.
Most of the front-facing surfaces were additionally covered with black tape to reduce potential reflections.
The tape was removed for this photograph.}
\label{f:camera}
\end{figure}

The photocathode has an effective area with a diameter of ${\sim}25$\,mm, which is large enough to contain the spot and capture all of the potential HP signal emitted from the surface of the mirror, even in the presence of the seasonal movement of the spot~\cite{Dobrich:2014kda}.
The PMT was placed inside of a cooled housing FACT50~\cite{et-fact50} (see also \cref{f:camera}), which is able to reduce the internal temperature to 50\,K below the ambient temperature.
Nevertheless, for the measurements reported here, the cooling was disabled, mostly to avoid build-up of ice on the PMT pins and to remove the additional insulating double-pane window the housing had installed in front of the PMT.
Cosmic muons passing through this volume of glass could produce a significant quantity of Cherenkov photons, which would increase the levels of background.

\begin{figure}[t]
\centering
\includegraphics[width=\linewidth]{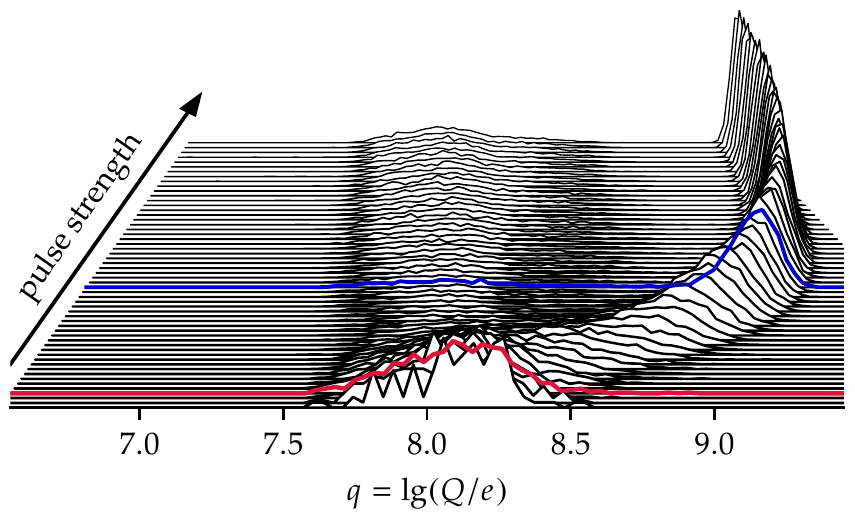}
\caption{A pseudo-3D plot of charge $Q$ collected at the anode of the PMT as a function of the intensity of the LED flasher (pulse strength in arbitrary units).
The red line indicates the setting used for the SPE study (see text for details) and the blue line is the higher setting used to acquire pulses shown in \cref{f:roi}.}
\label{f:flasher}
\end{figure}

The operating voltage of the PMT was set to 1050\,V which was determined from the observation of a break in the dark-count rate of a voltage scan~\cite{Dobrich:2015tpa}.
We then performed an operational test and a calibration of the PMT with a faint blue LED flasher emitting light pulses of adjustable strength.
In \cref{f:flasher}, we observe how the distributions of collected anode charge $Q$ (black lines) at particular pulse strengths develop a peak at large charges (i.e.\ at $Q>10^9$\,e) as the intensity of the LED is gradually increased and increasingly more photons are emitted in each pulse.
For the estimation of the SPE charge, the optimal setting of the LED was chosen such that \emph{(i)} 95\% of the captured traces ($4\,\upmu$s long) contained only one single pulse, \emph{(ii)} the PMT saw pulses only 20\% of the time, and \emph{(iii)} the pulse was observed in a narrow time interval ${\sim}290$\,ns after the flasher trigger.
From the charge distribution at this chosen LED intensity and high-voltage setting, we determined the gain of the PMT to be $G\approx10^8$.

\begin{figure}[t]
\centering
\includegraphics[width=\linewidth]{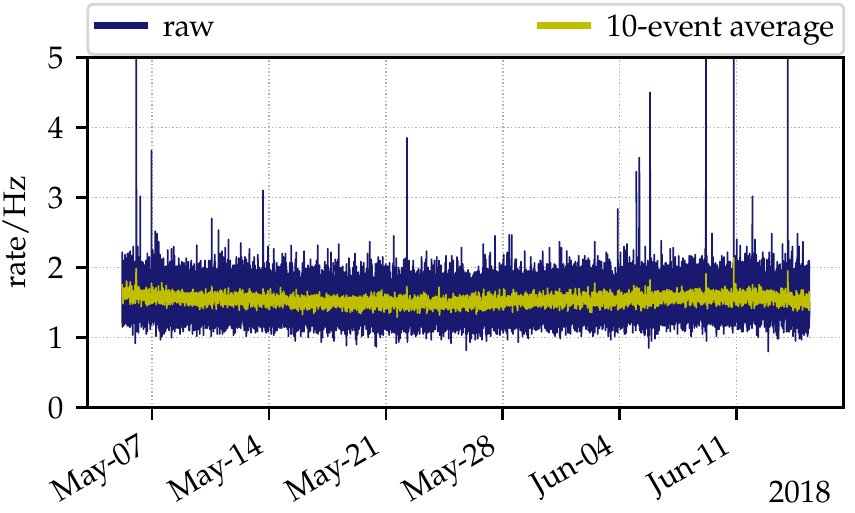}
\caption{Measurement of the internal background rate of the PMT.
Data were taken in 60\,s intervals, with the PMT inside the sealed FACT50 housing.}
\label{f:internal_background}
\end{figure}

The \emph{internal background} of the PMT was measured by attaching a light-tight metal lid onto the FACT50 housing, essentially isolating the PMT from all external light sources.
The trigger rates captured in this sealed-PMT condition for more than a month are shown in \cref{f:internal_background}, where a relatively stable and low trigger rate can be observed with a mean of 1.55\,Hz and a standard deviation of 0.20\,Hz.
The isolated peaks, seen at certain times, are most likely the result of cosmic-ray muons directly hitting the glass envelope of the PMT and thus releasing many Cherenkov photons.

\subsection{Mechanical parts}

\begin{figure}[t]
\centering
\includegraphics[width=\linewidth]{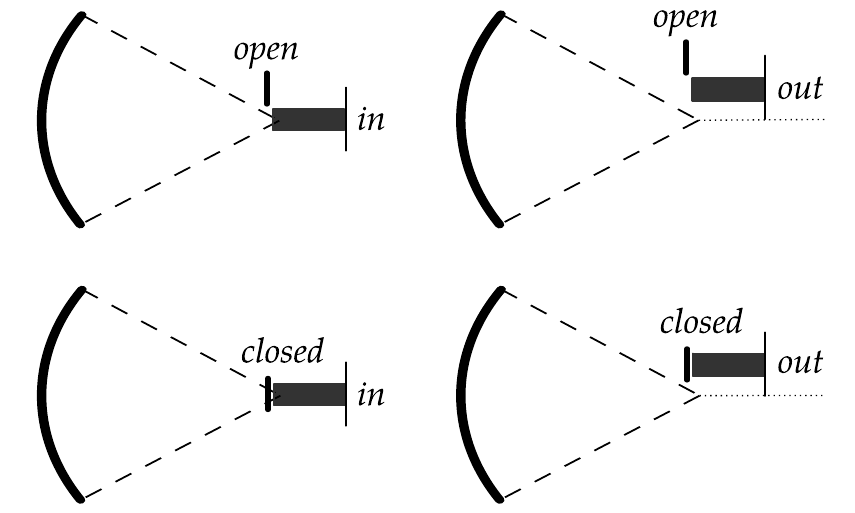}
\caption{Schematic drawing of the four modes in a measurement cycle.
For the two modes on the left, the PMT is placed \emph{in} the central point (CP) of the mirror and the shutter in front of the PMT is \emph{open} (top) and \emph{closed} (bottom).
The same shutter configurations are repeated with the PMT displaced by 71.5\,mm \emph{out} of the CP (right).
Simplified from~\cite{Experiment:2017icw,Experiment:2017kmm}.}
\label{f:configs}
\end{figure}

The PMT camera was placed in the CP and directed along the optical axis of the mirror, see \cref{f:setup,f:area}.
Our setup included motorized components that enabled us to obscure and move the PMT.
We installed a plate wrapped in black felt as a shutter in front of the PMT in a way that, when closed, it obscured the view of the PMT towards the mirror (see the drawing in \cref{f:configs}).
This shutter was operated by a Thorlabs MFF101 motorized flipper.
In addition to that, we placed the whole camera (PMT, FACT50 housing, and shutter) on an OWIS Precision Linear Stage LTM60-150-HSM with a total travel distance of 145\,mm.
The linear stage was installed horizontally and orthogonal to the optical axis of the mirror, enabling us to drive the PMT onto positions inside (\emph{in}) and outside (\emph{out}) of the CP, where the latter was at a distance of 71.5\,mm from the CP.
Since the experimental area is locked down during data acquisition, we had to control these mechanical components remotely.
For this task we chose a Raspberry Pi (mini)computer, which could be accessed and programmed remotely through a network connection.
As the PMT was exhibiting a non-negligible memory effect (see \cref{s:memory}) we decided to replace the Thorlabs shutter, which could obscure the PMT with only ${\lesssim}80\%$ efficiency, with a perfectly light-tight 6-blade-teflon iris-diaphragm Uniblitz NS65B with a bi-stable optical shutter with an aperture of 65\,mm.

Since the PMT is observing slightly different scenery behind the camera (through the reflection in the mirror) when it is moved in and out of the CP, we installed a felt baffle around the camera opening so that this view stays approximately the same in both positions, see \cref{f:camera}.

The standard measurement procedure we used was cycling between the positions \emph{out} of the CP (where no potential signal can be seen) and \emph{in} the CP (where we expect to see the potential DM signal).
Additionally, in both positions we measured with an \emph{opened} and \emph{closed} shutter.
This results in the following four modes per cycle: (a) \emph{out/open}, (b) \emph{out/closed}, (c) \emph{in/open}, and (d) \emph{in/closed}, as is schematically illustrated in \cref{f:configs}.
As will be discussed later in \cref{s:memory}, this cycling between the \emph{open} and \emph{closed} shutter also reduces the exposure of the PMT to the remaining ambient light and thus keeps the background rate of the PMT smaller.

\subsection{Data acquisition}
\label{s:daq}

The output current of the PMT was captured by a PicoScope 6404 digitizer with an 8-bit analog-to-digital converter (ADC). 
For the sampling period of the ADC, we choose $\Delta t=0.8$\,ns (i.e.\ sampling rate of 1.25\,GS/s), and we captured traces with $\pm1000$ samples around the trigger, thus storing traces with a total length of 1.6\,$\upmu$s.
The voltage range of the ADC was set to $\pm1$\,V at $50\,\Omega$ input impedance.
An internal trigger was set to a threshold of 8\,ADC counts (${\sim}63$\,mV) below a baseline (since PMT pulses are negative).
The ADC itself was then connected through a USB connection to a data-acquisition (DAQ) computer, which controlled the whole data-acquisition procedure.
The captured traces from the ADC were stored with additional data including the state of the measurement cycle (PMT position, shutter status), the PMT and ambient temperatures, air pressure etc.\ obtained from the slow control system.
The DAQ also controlled the movement of the mechanical components of the experiment.

The stored traces were analyzed off-line by running a baseline estimation algorithm first followed by a pulse-finding algorithm.
Most of the traces contained only one pulse that triggered the ADC; nevertheless, some traces contained additional pulses.
In the following, rates were obtained by counting all pulses in an event.

\begin{figure}[t]
\centering
\includegraphics[width=0.985\linewidth]{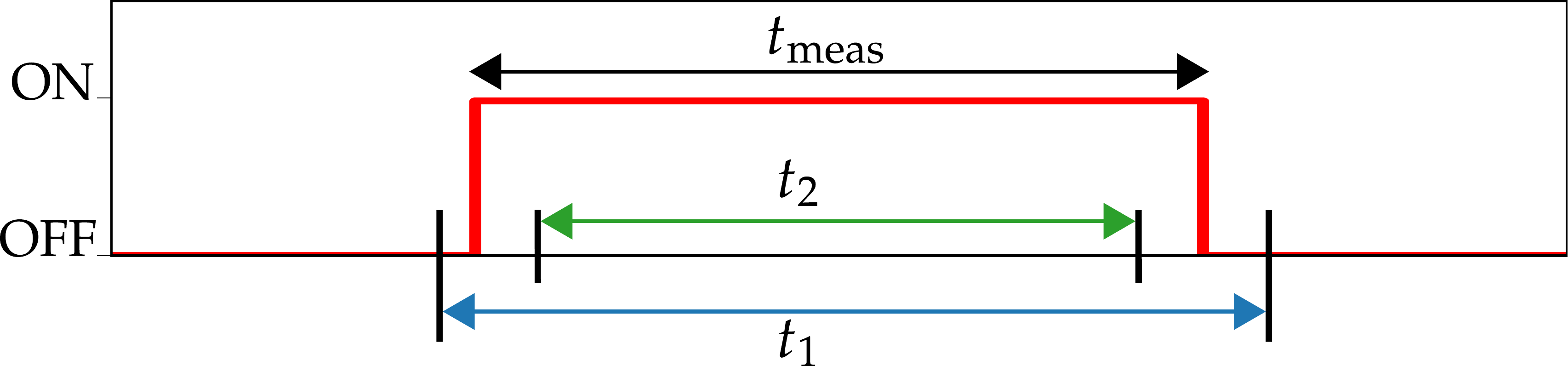}
\\[0.3cm]
\includegraphics[width=\linewidth]{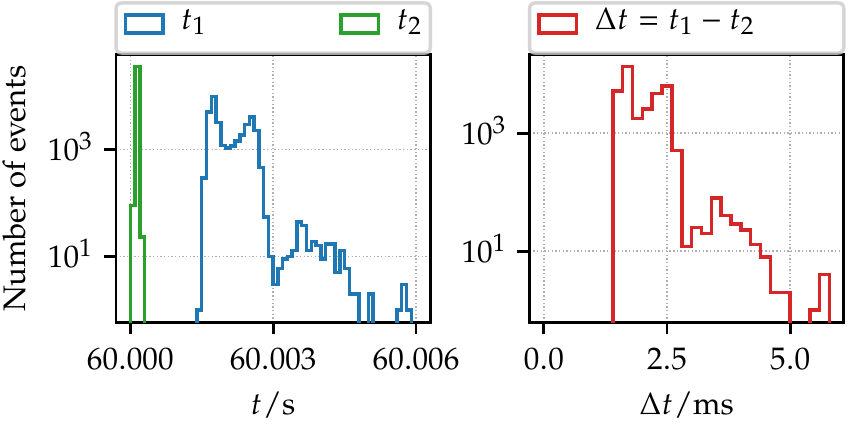}
\caption{
\emph{Top:} Schematic drawing of the implementation of the two timers within the DAQ procedure.
\emph{Bottom:} Distributions of the two timers $t_1$ and $t_2$ (left) and their difference $\Delta t$ (right).
Since $t_1$ includes the calls to the start and stop functions, its distribution is more spread out than that of $t_2$ which only measures the 60\,s of the sleep function. 
Note that the difference $\Delta t$ is shown in milliseconds.}
\label{f:timing_accuracy}
\end{figure}

\paragraph{Timing accuracy.}
The data acquisition of our experiment was split into events, each with a duration of 60 seconds.
With typical trigger rates of several Hz, this duration was short enough to have all the triggered traces stored in the internal memory of the ADC so that the DAQ would not be interrupted by trace transfers over the relatively slow USB connection.
In this way, no irregular dead time was introduced, and all the traces were read-out only at the end of the event duration.
Since the time delay between the call of the acquisition start function and the actual begin of the data acquisition by the ADC was not known, we implemented two timers within our DAQ software to obtain the smallest possible uncertainty on the exact duration of the measurement intervals (the \emph{on} time).
The first timer $t_1$ starts right before the call to the start function and ends right after the call to the stop function.
The second timer $t_2$ is started immediately after the call to the start function and stops right before the call to the stop function.
The true measuring time $t_\text{meas}$ thus lies somewhere between the two timer readings $t_2\leqslant t_\text{meas}\leqslant t_1$, as illustrated in \cref{f:timing_accuracy}-(top).

The distributions of the two times as well as their difference are shown in \cref{f:timing_accuracy}-(bottom).
The duration of an event is estimated to be close to the mean of both timers, $t_\text{event}\approx(t_1+t_2)/2$, and its uncertainty on the order of their difference, $\Delta t=t_1-t_2$.

For the particular run with $1.44{\times}10^7$ events, used for the analysis below, the mean event duration was $\langle t_\text{event}\rangle=60.00138$\,s, the mean time difference (and thus the uncertainty of the mean event duration) was $\langle\Delta t\rangle=0.00237$\,s, resulting in a relative time uncertainty of $\langle\Delta t\rangle/\langle t_\text{event}\rangle=4{\times}10^{-5}$.

\paragraph{Comparison of digitizers.}
For the purposes of systematic cross-checks of our DAQ, a different digitizer was employed in addition to the PicoScope described above.
The second digitizer (CAEN DT5751) had a higher ADC resolution of 10\,bits and a fixed input range of 1\,V at $50\,\Omega$ impedance.
In a special run, we installed the digitizers in parallel so that they both triggered on the same output of the PMT (via an impedance-matched signal splitter), and the trigger thresholds for both devices were set to equal levels.
Furthermore, in this run a rubber-sealed metal lid was used to completely obscure the PMT.
With this setup, we determined that the mean difference in the relative rate is ${\sim}(3.05 \pm 0.04)\%$.

\subsection{Muon-monitoring system}

\begin{figure}[t]
\centering
\includegraphics[width=\linewidth]{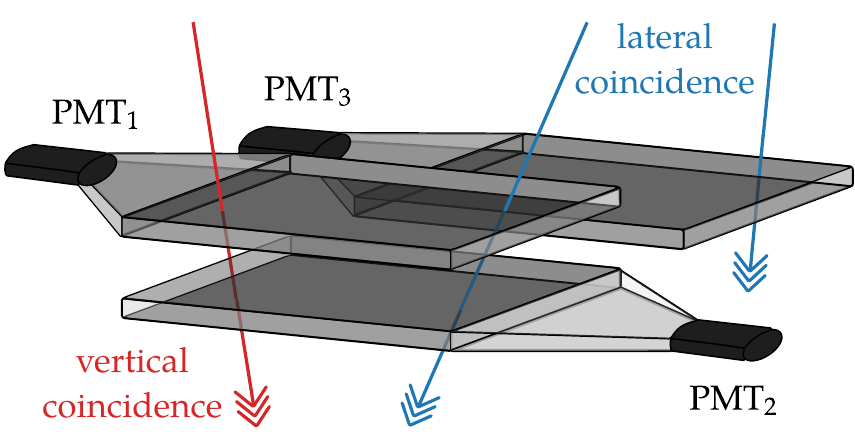}
\\[0.3cm]
\includegraphics[width=\linewidth]{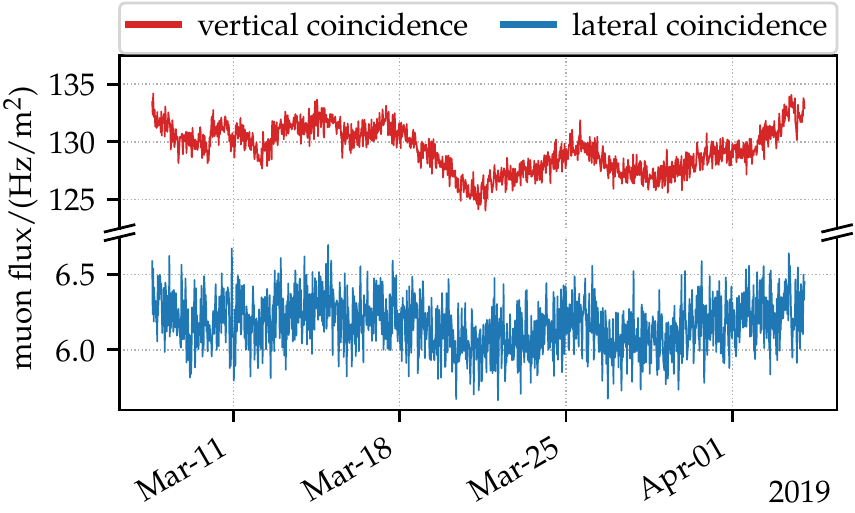}
\caption{\emph{Top:} Schematic diagram of the muon-monitoring system.
The system counts vertical coincidences of ionizing particles (red), when PMT$_1$ and PMT$_2$ are triggered, and lateral coincidence (blue), when PMT$_2$ and PMT$_3$ are triggered.
\emph{Bottom:} Measured mean rates for the vertical (red) and lateral (blue) coincidences.}
\label{f:muon_system}
\end{figure}

Since cosmic muons can easily penetrate thick layers of rock, they also reach our experimental hall through its ${\sim}2$\,m of concrete.
They therefore contribute to our background by emitting Cherenkov photons either in the glass enclosure of the PMT or in the air surrounding the experiment.
In order to systematically monitor variations in this potential source of background, we built a dedicated muon-monitoring system which was placed in a corner of the experimental hall.
The system consists of three scintillator bars ($\unit[62.5]{cm}\times\unit[25]{cm}\times\unit[2.2]{cm}$), each of which is connected to a PMT via a light guide.
The outputs of the PMTs are connected to discriminators, sending triggers to coincidence units which are in turn connected to counters, read out periodically by a slow-control computer (Raspberry Pi).
The scintillators and coincidence units were arranged so that we could monitor the temporal and spatial coincidence of muons, as schematically depicted in \cref{f:muon_system}-(top).
The captured coincidence count rates of our muon-monitoring system for a time period of approximately one year are shown in \cref{f:muon_system}-(bottom).
The typical vertical-coincidence rate for the muon monitor was ${\sim}\unit[130]{Hz/m^2}$.
Note that the muon monitor was not deployed as any kind of a veto system, nevertheless, in the next Section this coincidence data was used for correlation studies with measured rates to search for potential sources of background.

\section{Analysis}
\label{s:analysis}

We reported some preliminary results from previous measurement runs~\cite{Experiment:2017kmm,Veberic:2017icw}.
For the analysis discussed here, we use run number~v35, which was taken from March~07 to April~04,~2019 (with a duration of \unit[27.5]{days}).
To minimize eventual systematic drifts due to changing environmental conditions, after 60\,s of data taking we switch between each of the modalities in the measurement cycle shown in \cref{f:configs}.
This corresponds to an effective life-time of $\unit[4\times145.3]{hours}$ for the entire duration of run~v35.

\subsection{Sources of background}
\label{s:background}

\begin{figure}[t]
\centering
\includegraphics[width=\linewidth]{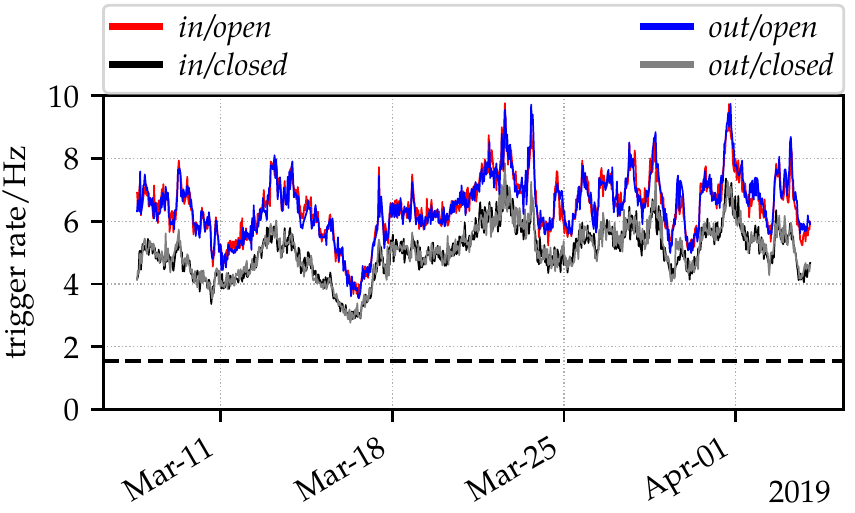}
\caption{Trigger rates recorded in run v35.
For this plot, the rates are averaged over 10 cycles (i.e.\ over $44$\,min).
The dashed line indicates the level of the internal dark-count rate (with the sealed PMT, see~\cref{f:internal_background}).}
\label{f:pmt_rates}
\end{figure}

\cref{f:pmt_rates} shows the averaged trigger rates in this dataset for the PMT operating in the four possible modes.
It is interesting to note that the rate observed with the \emph{closed} shutter is much higher than the dark-count rate (shown in \cref{f:internal_background}) and follows the variations of the rates with the \emph{open} shutter rather closely.
The origin of this phenomenon is discussed in \cref{s:memory}.
In the following, we make a distinction between internal and external sources of the background, based on whether their origins are the internal parts of the PMT or external factors.

As is clear from \cref{f:configs}, a positive difference between the \emph{open} rates, $\Delta r_\emph{open}=r_\emph{in/open}-r_\emph{out/open}$, is a proxy for the HP signal.
We found that this difference is uncorrelated with the internal background as measured in the \emph{closed} modes, $\Delta r_\emph{closed}=r_\emph{in/closed}-r_\emph{out/closed}$, implying that the potential detection of a HP signal is mostly limited by external nuisances.
Moreover, as can be seen in \cref{f:pmt_rates}, the typical rates are 3 to 4 times higher than the internal-background rate measured with the sealed PMT (see \cref{f:internal_background}).

In the following, we consider three main contributions to the dark-count rate, which originate from thermal and cosmic-ray-induced backgrounds, and summarize the conservative estimates of the size of these effects.
Rather than attempting to give an absolute size of each of the effects, we compare the relative variation of the rate, $v=\sigma(r)/\langle r\rangle$, with the relative variations of the background sources in question.

In the two \emph{out} modes, where no HP signal is expected, we observe relative variations on the order of $v_\emph{out/open}\approx20\%$ and $v_\emph{out/closed}\approx23\%$.
In the following, we will consider three possible sources of the background.
Nevertheless, none of them (nor their combination) can account for the variability observed in our data.

\paragraph{Thermal emission from the PMT photocathode.}
The rate of thermal emission of our PMT is estimated by rescaling existing measurements of the thermal response of bialkali photocathodes~\cite{Coates:1972}, in order to account for the difference in the photocathode area and collection efficiency.
Assuming that the dark-count rate observed in the sealed PMT is mostly of thermal origin, we obtain a linear dependence of the rate on the temperature with an average gradient of 0.07\,Hz/K, albeit only within the range of temperatures encountered during the measurement ($10^\circ$ to $20^\circ$C).
We monitored the standard deviation of the temperature to 0.2\,K, over the entire run, which ultimately yields $v_\text{temp}\lesssim1\%$ for the dark-count rate due to thermal emission.

\paragraph{Thermal emission from the environment.}
For simplicity we assume that the PMT in the \emph{open} mode, which has a field of view of $2\pi$\,sr towards the mirror, observes an isotropic blackbody-like source with a temperature of 300\,K.
The rate of photons from such a thermal source, that possess wavelengths within the range where the PMT has non-negligible quantum efficiency, is on the order of $10^{-7}$\,Hz. Clearly, this is insufficient to explain the observed background.
Furthermore, the quantum efficiency of the PMT decreases rapidly in the infrared wavelengths, where the thermal Planck spectrum at room temperatures actually starts to rise.

\paragraph{Cherenkov photons from cosmic-ray muons.}
The threshold energy of muons for production of Cherenkov photons in air under normal conditions ($T=15^\circ$C, $P=1013.25$\,hPa, dry) is ${\sim}4$\,GeV.
The energy loss of muons in the concrete shielding of the building is estimated to be ${\sim}1$\,GeV.
The muon monitor will thus be triggered by muons with energies above ${\sim}1$\,GeV, while the relevant contribution to the Cherenkov background comes from the muons with energies above ${\sim}5$\,GeV (before the concrete shielding).
The vertical muon intensities for the two energies are $\mathrm{d}N/\mathrm{d}A\,\mathrm{d}\Omega\,\mathrm{d}t|_{\theta=0^\circ}=73$ and 23\,/\,m$^2$\,sr\,s, respectively, the former being compatible with the average vertical-coincidence rate observed in our muon monitors.
We estimate the rate of associated Cherenkov photons with wavelengths between 150 and 640\,nm as around 90\,photons\,/\,m of muon travel in air and 120\,photons\,/\,mm in the fused-silica window of the PMT.
This corresponds roughly to a flux density of ${\sim}3000$\,photons\,/\,s\,m$^3$ being emitted within the experimental area (see~\cite{Badino:2000} for comparisons).
Direct muon hits can be identified in the event selection, as will be discussed later.
Since the rate of Cherenkov photons is directly proportional to the muon rate, we estimate the relative variation of the Cherenkov background $v_\text{ch}$ from the relative variation of the muon rate, $v_\mu\approx2.6\%$ as estimated from the muon monitoring.
Nevertheless, this variation is an order of magnitude smaller than that observed in the data.
Therefore, the background cannot be predominantly driven by the Cherenkov photons produced by cosmic muons.

\subsection{Volume effect}
\label{s:volume_effect}

\begin{figure}[t]
\centering
\includegraphics[width=\linewidth]{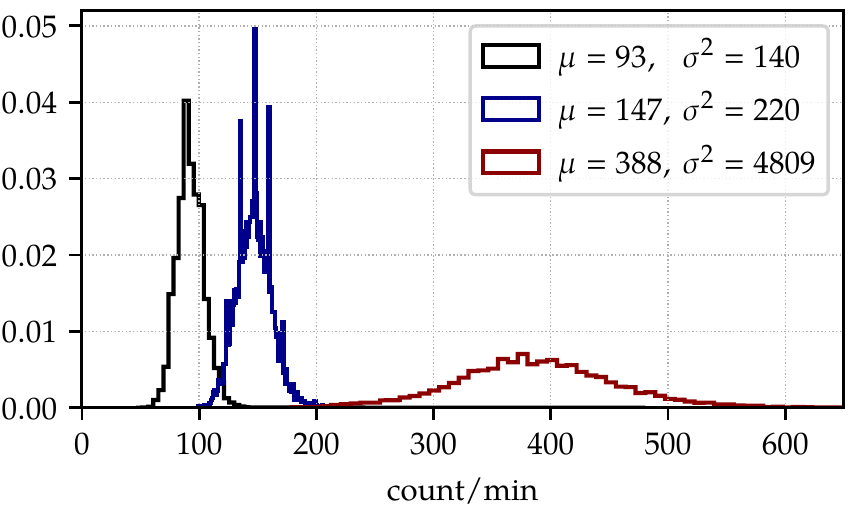}
\caption{Distribution of the trigger rates obtained during the measurements of the dark counts with the sealed PMT (black), the dark counts in the reduced volume box (blue), and counts during the regular run (dark red).
For the last two, the data were taken from the measurement with an \emph{open} shutter.
The inset shows their mean rates $\mu$ and variances $\sigma^2$.
Note that for independent counts (homogeneous Poissonian process) $\sigma^2$ should be close to $\mu$.}
\label{f:raw_triggers_distn}
\end{figure}

The PMT is set up within a light-tight experimental area with an effective volume of ${\sim}100$\,m$^3$.
In the previous section, we argued that our background fluctuations are mostly driven by external sources, although we could not identify a strong correlation with a particular source of background.
We thus investigated the dependence of the dark-count rate on the volume surrounding the PMT, motivated by the fact that the amount of detected Cherenkov light in underground caves scales with their dimensions~\cite{Badino:2000}.

We performed a series of dark-count measurements with a progressively smaller light-tight box built around the PMT while operating the shutter in the same way as for a regular run.
Each of these runs lasted for 2~days.
Initially, the box had a volume ${\sim}500{\times}$ smaller than the experimental area.
This volume was then increasingly reduced to ${\sim}2500{\times}$ smaller than the experimental area.
As an immediate effect, we observed a significant reduction of the trigger rates and its clear proportionality to the volume observed by the PMT.

The distribution of the trigger rates recorded with the installed box is shown in \cref{f:raw_triggers_distn} and compared to the results from run v35 and the internal-background rate measured with the sealed PMT.

This leads us to the conclusion that our background is, to some extent, dependent on the effective volume of the region surrounding or observed by the PMT.
Nevertheless, since we already rejected Cherenkov or thermal photons as our main background, we can at this point only speculate about other possible sources~\cite{helbing2003light,Aartsen:2016nxy}.

\subsection{Event selection}
\label{s:pulse_selection}

\begin{figure}[t]
\centering
\includegraphics[width=\linewidth]{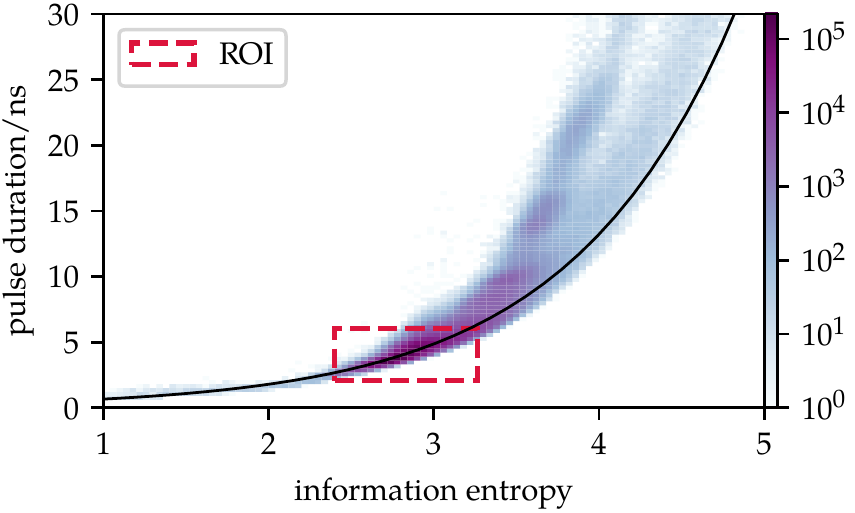}
\\[0.3cm]
\includegraphics[width=\linewidth]{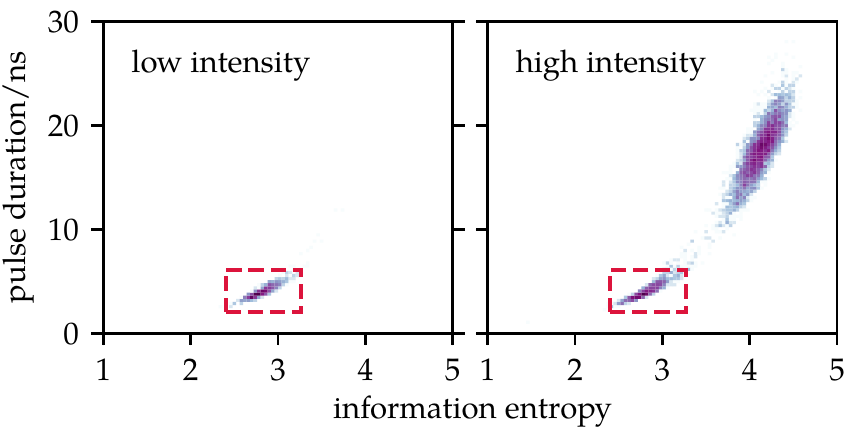}
\\[0.3cm]
\includegraphics[width=\linewidth]{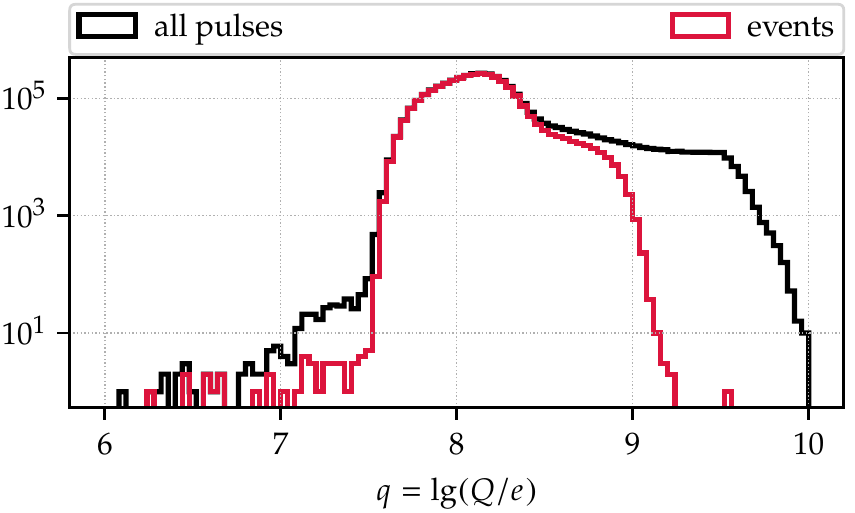}
\caption{\emph{Top:} Distribution of the information entropy and duration of recorded pulses.
The black line indicates the relation between these two parameters for pulses with a Gaussian shape in time.
The red-dashed line indicates the SPE-like selection applied to the pulses (ROI).
\emph{Middle:} Same as above but for the two settings of the LED-flasher strength. The information entropy for the weak strength, which produces mostly SPE-like pulses, is shown on the left, while the same for the higher setting producing pulses with multiple photons is shown on the right, illustrating the particular choice for ROI above.
\emph{Bottom:} Distribution of the pulse charge $Q$ before and after event selection.}
\label{f:roi}
\end{figure}

The data acquisition is performed in a self-triggering mode, as described in \cref{s:daq}, whereas pulse finding in recorded traces and subsequent event-building is performed with an off-line algorithm.
The HP signal is expected to appear as an increased rate of SPE-like pulses (in the \emph{in/open} mode). Therefore, a mild event selection is applied to achieve the best possible single-photon counting accuracy.

Pulses are selected according to the following two properties: duration of the pulse and its information (Shannon) entropy.
These properties were chosen to \emph{(i)} avoid direct selection based on the pulse amplitude or charge (since they both have standard deviations comparable to their means, and thus, have poor discrimination power) and \emph{(ii)} suppress long-duration pulses which are most likely associated with photon bursts from non-HP sources.

The chosen selection criteria define a region of interest (ROI), the parameters of which were determined and refined through a series of calibrations with the single-photoelectron dataset obtained with an adjustable blue-LED source (red line in \cref{f:flasher}).
The ROI and charge distributions are shown in \cref{f:roi}.
Note that the ROI contains ${\sim}90\%$ of all the pulses, which we refer to as \emph{events} from now on.

\subsection{Deviations from homogeneous Poissonian process}
\label{s:non-poisson}

The recorded events exhibit two distinct features which deviate from standard Poisson statistics: \emph{(i)} an overdispersion (i.e.\ the variance is many times larger than the mean), and \emph{(ii)} the existence of correlations on short-time scales.

\begin{figure}[t]
\centering
\includegraphics[width=\linewidth]{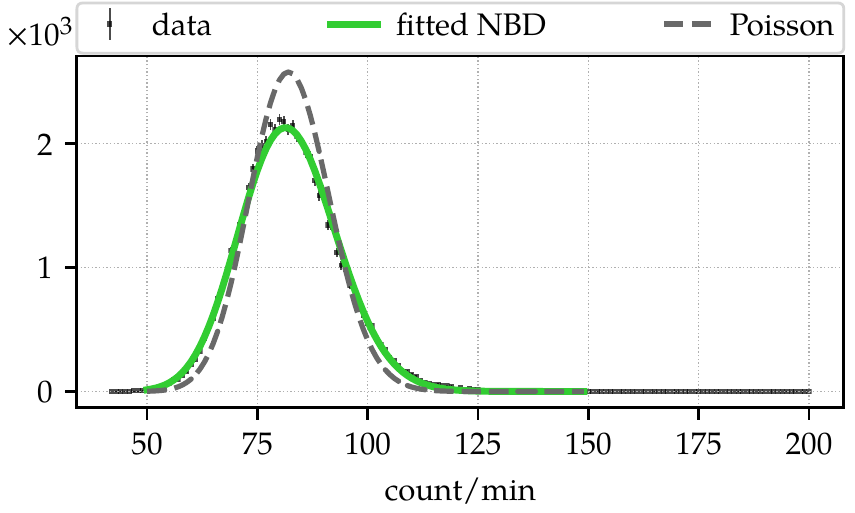}
\\[0.3cm]
\includegraphics[width=\linewidth]{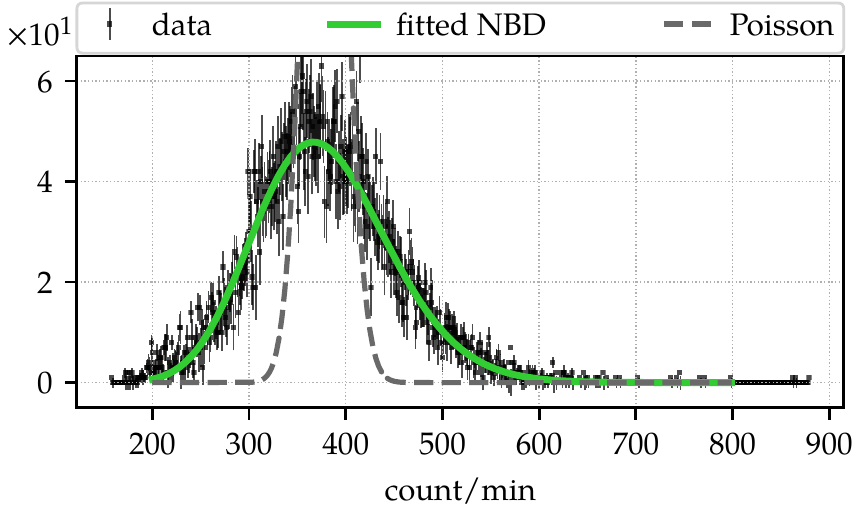}
\caption{\emph{Top:} Distribution of event counts in the internal-background run with the sealed PMT.
The distribution is quite close to the expected homogeneous Poissonian (dashed gray) and the negative-binomial (NBD) fit is slightly better (green).
\emph{Bottom:} The same for the regular run.
Here, the homogeneous Poisson distribution is a much worse description than the NBD.}
\label{f:overdispersion}
\end{figure}

The overdispersion is illustrated in \cref{f:overdispersion} and can be successfully described by allowing for nonhomogeneity in the counting process, e.g.\ with the Gamma heterogeneity of the Poissonian rate~\cite{Molenberghs:2010}.
The resulting marginal distribution is a negative binomial (NBD) whose mean always exceeds the variance~\cite{boziev1993} and fits the observed distribution of rates rather well.

We investigated the distribution of the pulse arrival-times in an independent run during which the PMT was kept at a fixed position and the shutter remained \emph{open} (mode \emph{out/open}).
The data were captured in intervals of 60\,s and the complete measurement lasted for 10~days.
The recorded sample contains $4.95{\times}10^6$ pulses, which were tagged and processed off-line.
The pulses were then selected according to the SPE-like criteria discussed in \cref{s:pulse_selection}.

In a homogeneous Poissonian process with a mean rate $\lambda$, the distribution of the waiting time $t$ between two consecutive events is exponential, $\mathrm{d}N/\mathrm{d}t=\lambda\exp(-\lambda t)$.
A useful transformation of the waiting times, $\Theta(t)=1-\exp(-\lambda t)$, maps short and large times towards 0 and 1, respectively, to give us a uniform distribution $\mathrm{d}N/\mathrm{d}\Theta=1$ for $\Theta$ in the half-open interval $[0, 1)$.
We can use the deviation from this uniform distribution as a suitable measure of heterogeneity of the arrival process.

\begin{figure}[t]
\centering
\includegraphics[width=\linewidth]{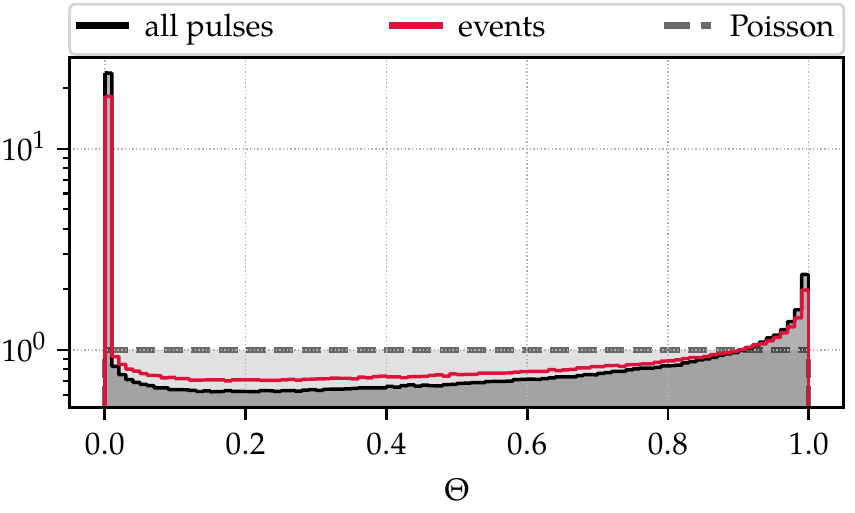}
\\[0.3cm]
\includegraphics[width=\linewidth]{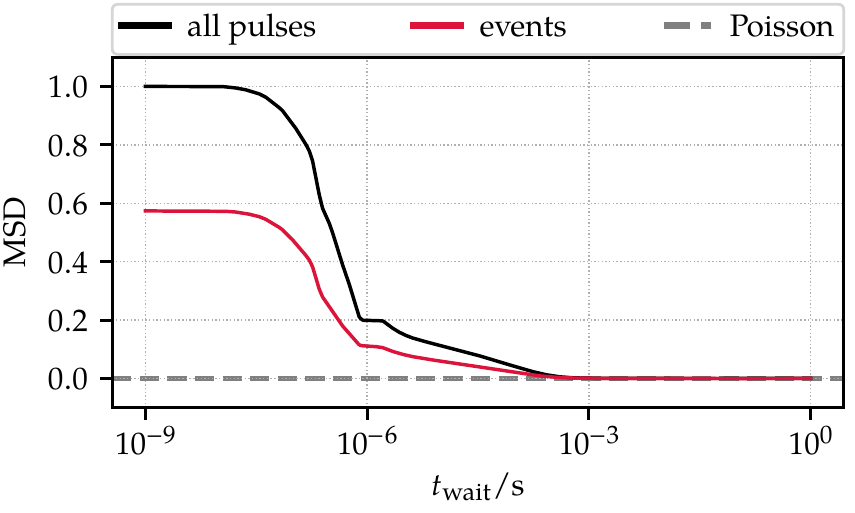}
\\[0.3cm]
\includegraphics[width=\linewidth]{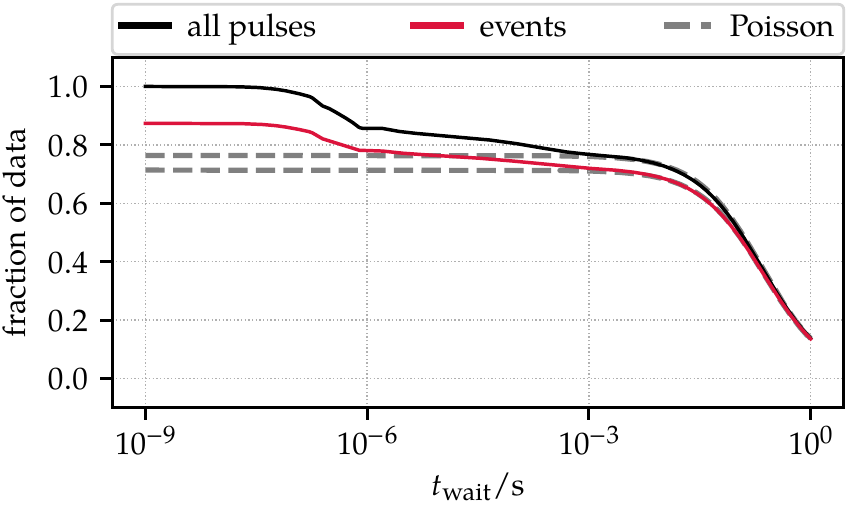}
\caption{The dashed gray line in all the plots above represents a corresponding homogeneous Poissonian reference, deviations from which illustrate the heterogeneity of the process.
\emph{Top:} Distribution of the $\Theta$-transformed waiting times (see text for more details).
\emph{Middle:} Mean-squared deviation (MSD) of the $\Theta$-transformed distribution (top) to a uniform distribution (homogeneous Poissonian) as function of $t_\text{wait}$.
\emph{Bottom:} Remaining fraction of data as a function of $t_\text{wait}$ cut. See \cref{f:roi} for the definition of ``events''.}
\label{f:counting}
\end{figure}

In \cref{f:counting}-(top), we plot the distribution of the $\Theta$-transformed intervals between successive events, referred to as waiting times.
The $\Theta$ transformation clearly reveals the deviation from the homogeneous Poisson process, especially the substantial clustering of the time-correlated events with small waiting times, either by PMT after-pulses or photon bursts.

To investigate the time scales on which the clustering occurs, we apply a waiting-time selection where consecutive events closer than $t<t_\text{wait}$ are counted only as one.
Selected waiting times from a homogeneous Poisson process can also be transformed to a uniform distribution with the transformation $\Theta'(t)=1-\exp[-\lambda(t-t_\text{wait})]$.

In \cref{f:counting}-(middle), we plot the mean-squared deviation (MSD) of the uniform distribution and the distribution of $\Theta'$-transformed waiting times as a function of $t_\text{wait}$.
The MSD gives a value of zero for a uniform distribution and is normalized such that the maximum possible deviation gives a value of 1.
We observe two significant departure points from homogeneous Poisson statistics.
The first break occurs on a millisecond scale ($10^{-3}$\,s) and is most likely related to cosmic-ray showers.
The second break appears on a microsecond scale ($10^{-6}$\,s) and is probably related to PMT after-pulsing since this can occur within a time window of 100\,ns to a few $\upmu$s.
Nevertheless, for our PMT, after-pulses normally amount to only 1\% of the count rate~\cite{et-manual}.
In \cref{f:counting}-(bottom), we show the efficiency of the waiting-time selection as a function of $t_\text{wait}$.
Roughly 23\% of events are clustered on scales below 1\,ms; 14\%, below $1\,\upmu$s.

We do not expect that HPs can induce such time-correlated events.
Nonetheless, in order to derive a conservative limit in \cref{s:results}, we do not perform any waiting-time selection or any other rejection based on the arrival time of the pulses.

\subsection{Memory effect}
\label{s:memory}

When comparing the sealed-PMT rates (\cref{f:internal_background}) with the \emph{closed} rates during the measurement run (\cref{f:pmt_rates}), we immediately see that the latter are two to five times higher.
Furthermore, the \emph{closed} rates in \cref{f:pmt_rates} also tightly follow the \emph{open} ones, indicating a certain level of correlation between the \emph{open} and \emph{closed} measurements.
Additionally, we also noticed that after each visit to the experimental area, which unavoidably brings a certain level of illumination, the PMT needed several hours to reduce and stabilize its rate.
To quantify the dependence of the rate on the historical levels of illumination, we devised a special run where the shutter was initially kept closed for 200\,min, after which it was open for a duration of 30\,min.
This sequence was then repeated several times.
Identically to the regular runs, we recorded the number of triggers in the same intervals of 60\,s.

\begin{figure}[t]
\centering
\includegraphics[width=\linewidth]{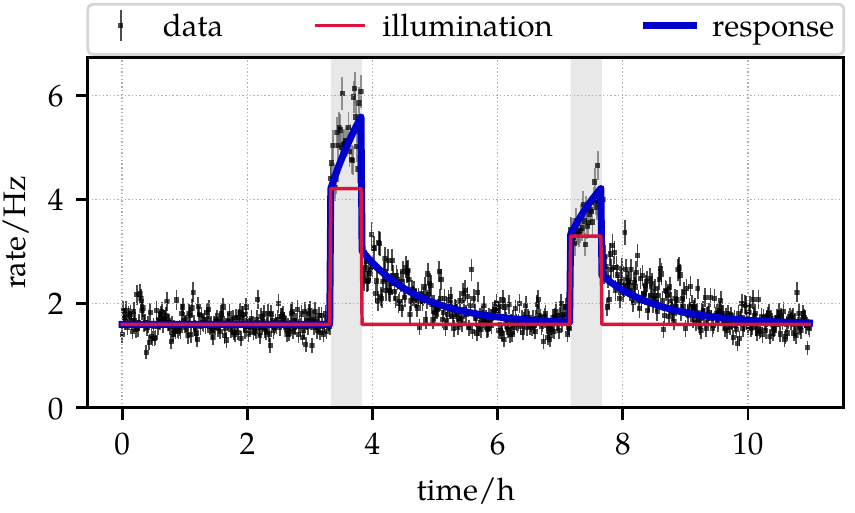}
\caption{An example of the trigger rate (black points) measured with a sequence of 200\,min in \emph{closed} mode and 30\,min in \emph{open} mode.
The grey-shaded region indicates the times at which light came in.
The illumination model (red) is fitted to the data along with the detector response (blue), as described by \cref{e:memory}.}
\label{f:memory}
\end{figure}

This measurement is plotted in \cref{f:memory}.
After the initial fast rise, the photon rate continues to increase slowly as the PMT photocathode continues to be exposed to the external sources of background light.
After the shutter is closed, the rate slowly decays, on the timescale of an hour, back to the typical value of the internal background as seen with the sealed PMT.
From this experiment, we conclude that in this particular low-noise PMT a \emph{memory} of the past history of the illumination exists.
It is striking that this effect extends over such long periods and even when the photocathode has been exposed only to a very faint external source of light (few Hz).

Based on these findings, we model the output response $r(t)$ of the PMT to an external input $s(t)$ as having an instantaneous and historical part,
\begin{equation}
  r(t) = s(t) + \alpha \int_{-\infty}^t \mathrm{d}t'\,K(t', t)\,s(t'),
\label{e:memory}
\end{equation}
where an exponential decay,
\begin{equation}
  K(t', t) = \frac1\tau \exp\left(-\frac{t-t'}\tau\right),
\label{e:memory_kernel}
\end{equation}
is chosen as the memory kernel.
The memory efficiency $\alpha$ and the decay time $\tau$ are characteristics of the PMT and are obtained from fits of the
data in \cref{f:memory} as
\begin{align}
\begin{split}
  \alpha &= 0.922 \pm 0.015,
\\
  \tau/\unit{min} &= 54.124 \pm 1.115.
\end{split}
\label{e:memory_param}
\end{align}

\begin{figure}[t]
\centering
\includegraphics[width=\linewidth]{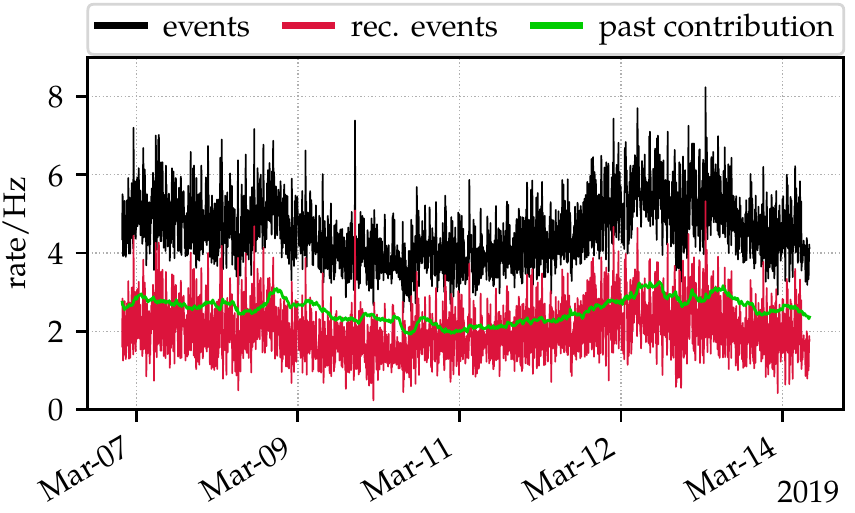}
\caption{An example of the event-rate reconstruction (red), showing a one-week sample plotting only the \emph{out/closed} data.
The measured event rate is shown in black and the contribution from the past illumination is shown in green.}
\label{f:reconstruction}
\end{figure}

In this model, the input illumination of the PMT thus influences output only up to several decay times $\tau$ into the future.
Ignoring for the moment the internal background, a constant input rate $s$ will therefore converge to an output rate $r=(1+\alpha)s$ in a time span of several $\tau$.
In our standard measurement cycle, where each mode lasts for $\Delta t = 60\,$s, this constant input $s$ is periodically blocked and unblocked by the shutter.
With such an input, the output will converge to the following \emph{open} and \emph{closed} expected rates at the middle of the cycle intervals,
\begin{align}
\begin{split}
r_\text{open} &=
  s \left[1 + \alpha \left(1 - \frac{f}{1+f^2}\right)\right],
\\
r_\text{closed} &=
  \alpha\,s \left(\frac{f}{1+f^2}\right),
\end{split}
\label{e:saturation_rate}
\end{align}
where we use the shorthand $f=\exp(-\Delta t/2\tau)$.
For $\Delta t\ll\tau$ this approximately simplifies to $r_\text{open}\approx s(1+\alpha/2)$ and $r_\text{closed}\approx\alpha s/2$ such that the difference is roughly proportional to the external input, $r_\text{open}-r_\text{closed}\approx s$ (c.f.\ \cref{f:pmt_rates}).

With these values known, we can fully correct for the historical memory effect to recover the instantaneous signal in our actual run.
Let us suppose that we start the data acquisition at $t_\circ=0$ and that the modes of the measurement cycle are changed at discrete intervals $t_i=i\Delta t$ for $i\geqslant0$.
Since $\tau$ is large compared to the sampling interval $\Delta t$, the average event-count over an $i$th interval $(t_{i-1},t_i]$ can be well approximated as in the middle of the interval, i.e.\ $\bar{t}_i=(t_{i-1}+t_i)/2$.
\cref{e:memory} can then be written for each time bin $i$ in terms of discrete sums as
\begin{equation}
\begin{split}
r_i =
  s_i & + \alpha \, e^{-\nicefrac{\bar{t}_i}{\tau}} \times
\\
  & \times
  \left[
    s_\circ +
    \textstyle{\sum}_{j=1}^{i-1}
      s_j (e^{\nicefrac{t_j}{\tau}} - e^{\nicefrac{t_{j-1}}{\tau}}) +
    s_i(e^{\nicefrac{\bar{t}_i}{\tau}} - e^{\nicefrac{t_i}{\tau}})
  \right],
\end{split}
\label{e:memory_sum}
\end{equation}
where $r_i=r(\bar{t}_i)$, $s_i=s(\bar{t}_i)$, and $s_\circ$ corresponds to the average input signal before the start of the measurement, $t<0$. 
Since this state is unknown, we use \cref{e:memory_sum} in fitting the PMT response data for the first few hours to obtain the best $s_\circ$.
Then the true signal can be iteratively reconstructed as
\begin{equation}
s_i =
  \frac{r_i - \alpha \, (h_i + s_\circ e^{-\nicefrac{\bar{t}_i}{\tau}})}
       {1 + \alpha \, (1 - f)},
\label{e:true_signal}
\end{equation}
where the auxiliary variable $h_i$ is obtained as
\begin{equation}
\begin{split}
h_\circ &= 0,
\\
h_i &=
  h_{i-1} \, f^2 + s_{i-1} \, f(1 - f^2),
  \qquad
  i\geqslant1.
\end{split}
\end{equation}

As is clear from \cref{e:true_signal}, any bias related to the exact value of $s_\circ$ can be avoided by simply discarding the first few hours of the reconstructed data.
Thereafter, given the parameters $\alpha$ and $\tau$, the reconstruction is close to exact.
An illustration of this procedure is shown in \cref{f:reconstruction}.

\subsection{Low-frequency background}
\label{s:detrending}

\begin{figure}[t]
\centering
\includegraphics[width=\linewidth]{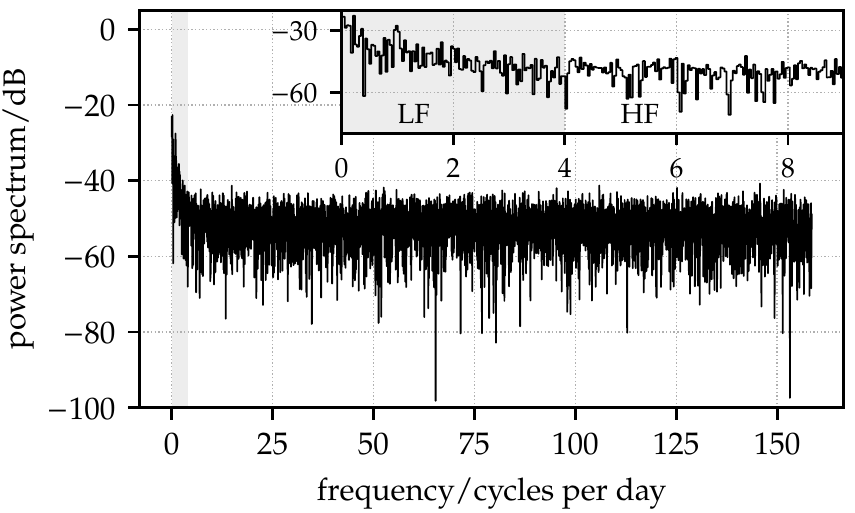}
\\[0.3cm]
\includegraphics[width=\linewidth]{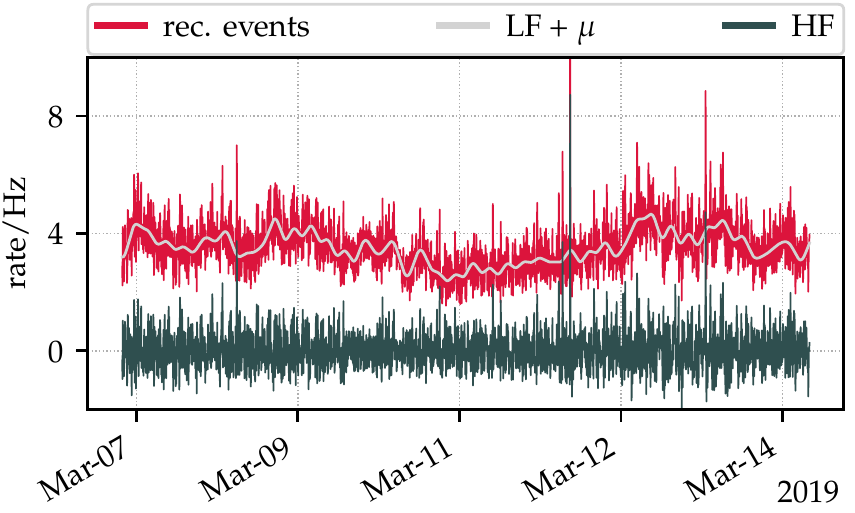}
\caption{\emph{Top:} Spectral Fourier decomposition of the reconstructed signal rate.
Frequency is plotted from the first harmonic up to the Nyquist frequency, and the corresponding spectral powers are normalized with respect to the stationary (constant) component.
The shaded region corresponds to the low-frequency (LF) truncation used in the bottom plot.
\emph{Bottom:} Detrending of the reconstructed rate as described in \cref{s:detrending}.
Data is taken from the \emph{out/open} mode of the measurement.
For plotting purposes, the LF harmonics have been shifted up by the mean event rate $\mu$.}
\label{f:time_modulation}
\end{figure}

A visualization of the event rate in the frequency domain (see \cref{f:time_modulation}) reveals a relatively significant modulation of the signal at low frequencies (i.e.\ from zero to several cycles per day).
Since these harmonics are observed also in the measurement with the closed shutter, they are treated as background.
Furthermore, for our aimed level of sensitivity (${\sim}10^{-3}$\,Hz or ${\sim}1$ detectable HP-to-photon conversion every 15\,min), we do not expect any substantial time modulation of the HP signal.

Modulations on larger time scales are most likely related to modulations in the cosmic-ray flux, which are in turn modulated by variations in atmospheric pressure.
In our analysis, we detrend the reconstructed dataset by removing these low-frequency (LF) oscillations up to a certain frequency threshold.
The threshold is chosen such that it does not affect the event distribution of the higher-frequency (HF) residual.
This procedure is illustrated in \cref{f:time_modulation}, where components of the Fourier spectrum with frequencies less than 4\,cycles/day have been removed.

\subsection{Systematic uncertainties}
\label{s:systematics}

\paragraph{Timing uncertainty.}
A systematic shift of the timing estimate $t_\text{meas}$ with the uncertainty quoted in \cref{s:daq} would result in a systematic bias of 0.004\% of the average count rate for a given measurement mode.
Given a typical rate of 4\,Hz for the \emph{open} mode, this implies a systematic uncertainty of $3{\times}10^{-4}$\,Hz for the \emph{in}--\emph{out} difference.

\begin{figure}[t]
\centering
\includegraphics[width=\linewidth]{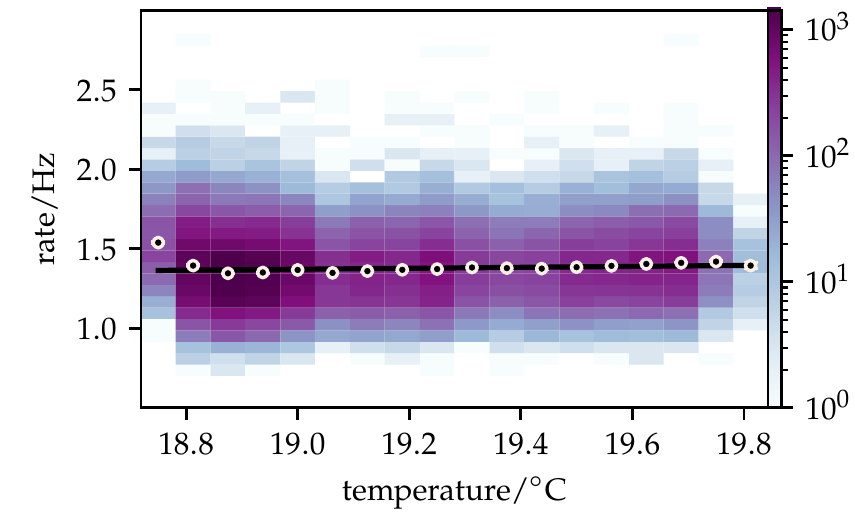}
\caption{Distribution of the internal-background event-rates in a sealed PMT as a function of the ambient temperature.
A linear fit yields a slope of $\unit[(0.030\pm0.014)]{Hz/^\circ C}$, which is compatible with our first estimation.}
\label{f:temperature_dependence}
\end{figure}

\paragraph{Temperature drift.}
The \funk experiment operates in environmental conditions with excellent temperature stability, as testified by the continuous monitoring before, during, and after the measurement runs.
Systematic errors can arise from fluctuations occurring between changes of the measurement configuration.
For our run v35, the typical variation of the temperature difference measured between the \emph{in} and \emph{out} positions is evaluated to be 0.02\,K.
Based only on general properties of the bialkali photocathodes, we estimated the temperature coefficient of the PMT rate to be around 0.07\,Hz/K in \cref{s:background}.
Here, we measure this coefficient from the internal-background data of the sealed PMT, as shown in \cref{f:temperature_dependence}.
Over the whole run, this temperature dependence induces a systematic uncertainty of $\unit[6{\times}10^{-4}]{Hz}$ in the count rate or 3\% in the \emph{in}$-$\emph{out} difference.

\paragraph{Pressure dependence of the Cherenkov rate.}
While considering the flux density of Cherenkov photons produced in air in~\cref{s:background}, we estimated that there are roughly 300\,kHz of photons produced within the experimental area.
By comparison, the rate difference \emph{open}--\emph{closed} amounts to ${\sim} 1\,$Hz, which corresponds to a fraction of $4{\times}10^{-6}$ of the estimated Cherenkov rate.
The latter varies with the air density and, therefore, with the ambient temperature and pressure.
Within the pressure range of interest, we estimate that this dependence behaves linearly with a gradient of 0.5\,kHz/hPa at $15^\circ$C.
From the pressure monitoring, this coefficient yields a typical variation of 30\,Hz or 0.01\% in the Cherenkov rate, which translates into a systematic uncertainty of $1.2{\times}10^{-4}$\,Hz for the measured rate difference between the two open modes in the cycle.

\begin{figure}[t]
\centering
\includegraphics[width=\linewidth]{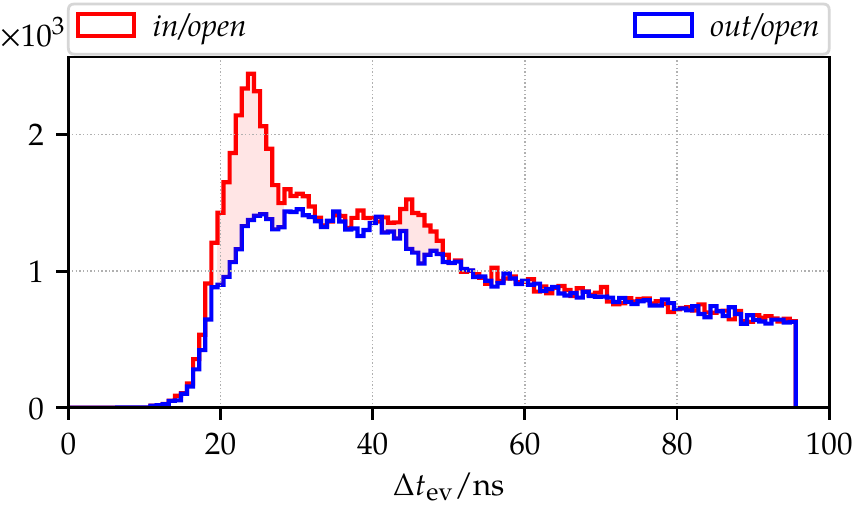}
\caption{Distribution of the time difference $\Delta t_\text{ev}$ between two consecutive events recorded within the same trace.
The red shaded region corresponds to the event excess due to the reflection off the mirror (see text for more details).}
\label{f:tp_diff}
\end{figure}

\paragraph{Mirror reflection.}
When the PMT is positioned at the center of the spherical mirror with the shutter open, some photons (e.g.\ from the muon hits) may be refracted towards the mirror and by design get perfectly reflected back to the PMT.
The magnitude of this effect has been evaluated by measuring time differences $\Delta t_\text{ev}$ between consecutive events and comparing them to the time a photon requires to cover the distance to the mirror and back (${\sim}22$\,ns).
In \cref{f:tp_diff}, we show an observation of two significant peaks that satisfy the timing requirement for single and double reflection (with a faint hint of a third).
The overall magnitude of this effect is estimated with a Monte-Carlo simulation, which gives a relative excess of $0.0188\pm0.0001$\,Hz for the event rate in the \emph{in/open} mode.
This excess, however, also includes contribution from the memory effect addressed in~\cref{s:memory}.
On account of the event-reconstruction efficiency (57\% for the \emph{open} data), we find a systematic bias of 0.01\,Hz for the \emph{in/open} mode.
We note that this is of the same order of magnitude as our statistical uncertainty for the difference \emph{in}$-$\emph{out}.

\begin{table}[t]
\centering
\caption{Summary of the systematic effects introduced by known sources and resulting in different measurement conditions at the positions \emph{in} and \emph{out}.}
\label{t:systematic}
\begin{tabular}{ll}
\toprule
source & $\sigma_\text{sys}/\text{Hz}$ \\
\midrule
DAQ measurement time & $3{\times}10^{-4}$ \\
PMT temperature drift & $6{\times}10^{-4}$ \\
Pressure dependence of Cherenkov production~~ & $1{\times}10^{-4}$ \\
Reflections from the mirror & $1{\times}10^{-2}$ \\
\bottomrule
\end{tabular}
\end{table}

\section{Results}
\label{s:results}

Assuming that the HP particles comprise the whole of the cold DM in the galactic halo, we can obtain our sensitivity to the kinetic-mixing parameter directly from \cref{e:power} as
\begin{align}
\chi &=
  4.1{\times}10^{-12}
  \left(
    \frac{r_\text{det}/q_\text{eff}}{\unit{Hz}} \,
    \frac{m_{\tilde\gamma}}{\unit{eV}}
  \right)^{1/2}
  \left(
    \frac{\eta \, A_\text{mirr}}{\unit{m^2}}
  \right)^{-1/2}
\nonumber
\\
  & \qquad\times
  \left(
    \frac{\langle\cos^2\theta\rangle}{2/3}
  \right)^{-1/2}
  \left(
    \frac{\unit[0.3]{GeV/cm^{3}}}{\rho_\text{CDM}}
  \right)^{1/2},
\end{align}
where $r_\text{det}$ is the DM-induced photon rate and $q_\text{eff}$ the quantum efficiency of the PMT.
The efficiency parameter $\eta=4(\sqrt{R}-R)^2/(1-R)^2$~\cite{arnaud_phd} accounts for the reflectivity $R$, where a value of $R=0.7$ is used as the reflectivity does not exceed this value for all wavelengths in the range of interest.
$\theta$ is the angle between the HP field polarization and the mirror plane.
The last two terms are arranged relative to the isotropic polarization distribution of the DM, where $\langle\cos^2\theta\rangle=2/3$, and $\rho_\text{CDM}$ is the local cold DM energy density.

The minimum detectable rate $r_\text{det}$ is governed by the dark-count rate and the total measurement time.
In our case, increasing the measurement time to one year would only improve the limit by a factor of 2.

Before proceeding to the derivation of a limit, we briefly summarize the three steps of the event selection procedure:
\begin{itemize}
\item[\emph{(i)}] individual pulses are tagged and events are selected by the SPE-like filters based on their duration and informational entropy (see \cref{s:pulse_selection}),

\item[\emph{(ii)}] the total event count for each 60\,s measurement mode is corrected for the memory effect, and the first 50~modes are discarded to avoid any potential bias (see \cref{s:memory}),

\item[\emph{(iii)}] the reconstructed rate is detrended in the frequency domain by removing Fourier components for frequencies of less than 4\,cycles/day (see \cref{s:detrending}).
\end{itemize}

\begin{figure}[t]
\centering
\includegraphics[width=\linewidth]{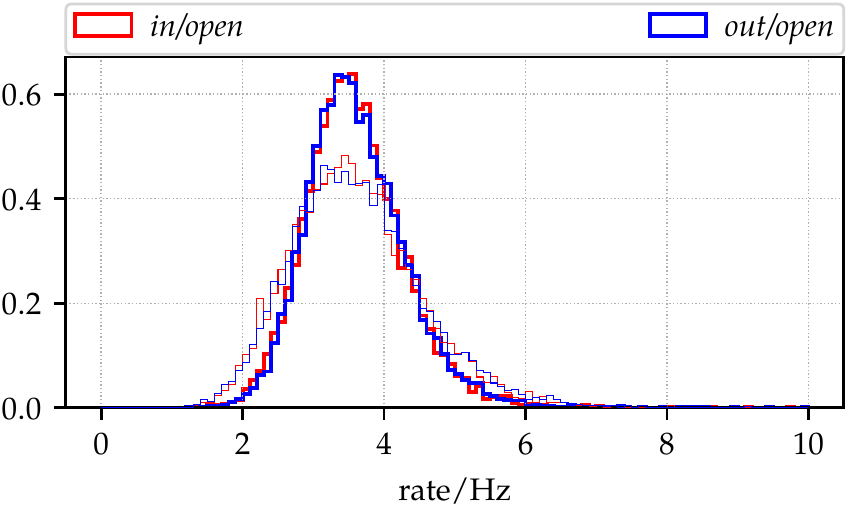}
\caption{Distribution of the reconstructed rates \emph{in} and \emph{out} of the radius point with the shutter \emph{open}.
The light and thicker lines correspond to the dataset before and after the detrending, respectively.}
\label{f:signal_distn}
\end{figure}

The rates after this selection are presented in \cref{f:signal_distn,t:rates}.
The potential HP ``signal'' is resolved from the difference of single-photon counts \emph{in} and \emph{out} of the CP with the shutter \emph{open}.
For this difference, we find
\begin{equation}
  \Delta r_\text{open}/\unit{Hz} = -0.0229 \pm 0.0108.
\label{e:signal_value}
\end{equation}
Furthermore, it is worth to mention that this negative signal is compatible with the difference we observe in the background-only measurement with the shutter \emph{closed},
\begin{equation}
  \Delta r_\text{closed}/\unit{Hz} = -0.0267 \pm 0.0106.
\end{equation}
We interpret $\Delta r_\text{closed}$ as a correctable systematic effect due to unknown sources and introduced by driving the PMT \emph{in} or \emph{out} of the CP.
Contributions from other known systematic sources are summarized in~\cref{t:systematic}.
The dominant effect comes from the mirror reflections $\Delta r_\text{refl}$, which we subtract from $\Delta r_\text{open}$.
The uncertainty on these reflections is evaluated to be less than 1\%, such that $\Delta r_\text{refl}=(1\pm0.01){\times}10^{-2}$\,Hz.
This gives an additional systematic uncertainty of $10^{-4}$\,Hz, which is added in quadrature with the uncertainties associated with the DAQ timing and temperature and pressure variations.
Our signal is therefore given by
\begin{align}
r_\text{signal}/\unit{Hz} &= \Delta r_\text{open} - \Delta r_\text{closed} - \Delta r_\text{refl}
\nonumber \\
&= -0.0062 \pm 0.0151\text{(stat)} \pm 0.0007\text{(sys)},
\end{align}
from which we conclude that there is no evidence for the presence of HP DM.
Let us remark that the rather large statistical uncertainty of $r_\text{signal}$ is a direct consequence of the overdispersion of counts (as discussed in \cref{s:non-poisson}), and that assuming homogeneous Poissonian statistics (as e.g.\ done by the Tokyo experiment~\cite{Suzuki:2015sza}) would result in a factor 4 smaller uncertainty and thus a factor 2 lower limit for $\chi$.

\begin{table}[t]
\centering
\caption{Rates measured in the four modes of the measurement.}
\label{t:rates}
\begin{tabular}{lll}
\toprule
$r$/Hz & \emph{in} & \emph{out} \\
\midrule
\emph{open} & $3.5937 \pm 0.0077 $ & $3.6166\pm 0.0078$ \\
\emph{closed}~~~~ & $2.2034 \pm 0.0074$~~ & $2.2301 \pm 0.0077$ \\
\bottomrule
\end{tabular}
\end{table}

\begin{figure}[t]
\centering
\includegraphics[width=\linewidth]{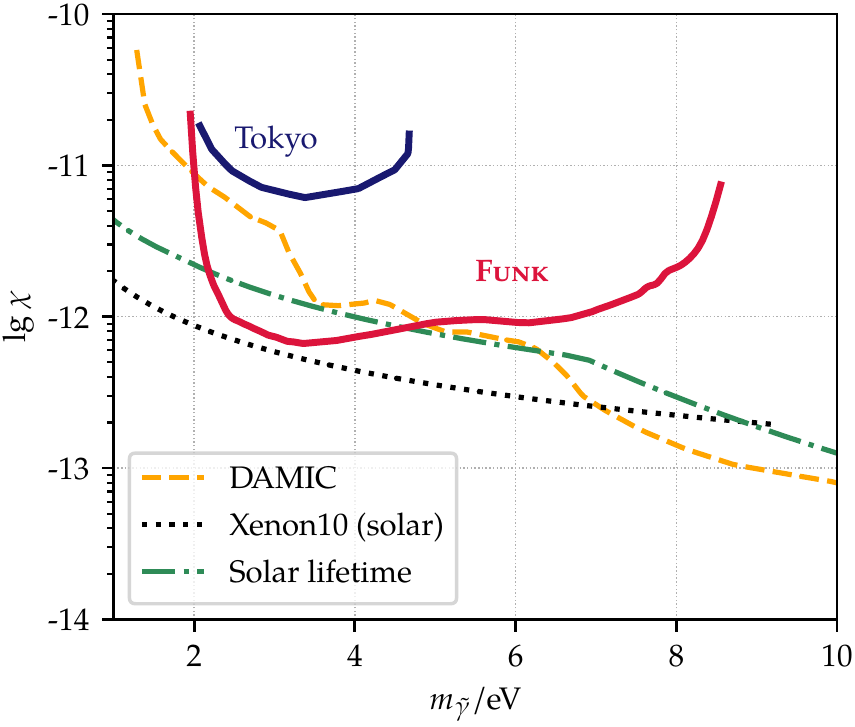}
\caption{
\funk{} exclusion limit at a 95\%~C.L. (red line). The limit labeled Tokyo~\cite{Suzuki:2015sza} was obtained by another dish antenna experiment using a smaller mirror.
The dashed orange line is a constraint from DAMIC~\cite{Aguilar-Arevalo:2019wdi}, searching for HP ionization signal in WIMP detector targets.
The dot-dashed green line corresponds to astrophysical bounds from the solar lifetime~\cite{Redondo:2013lna}, and the dotted black line is the limit from Xenon10~\cite{Bloch:2016sjj} for model of solar HPs.}
\label{f:limit_zoom}
\end{figure}

\begin{figure*}[t]
\centering
\includegraphics[width=0.85\textwidth]{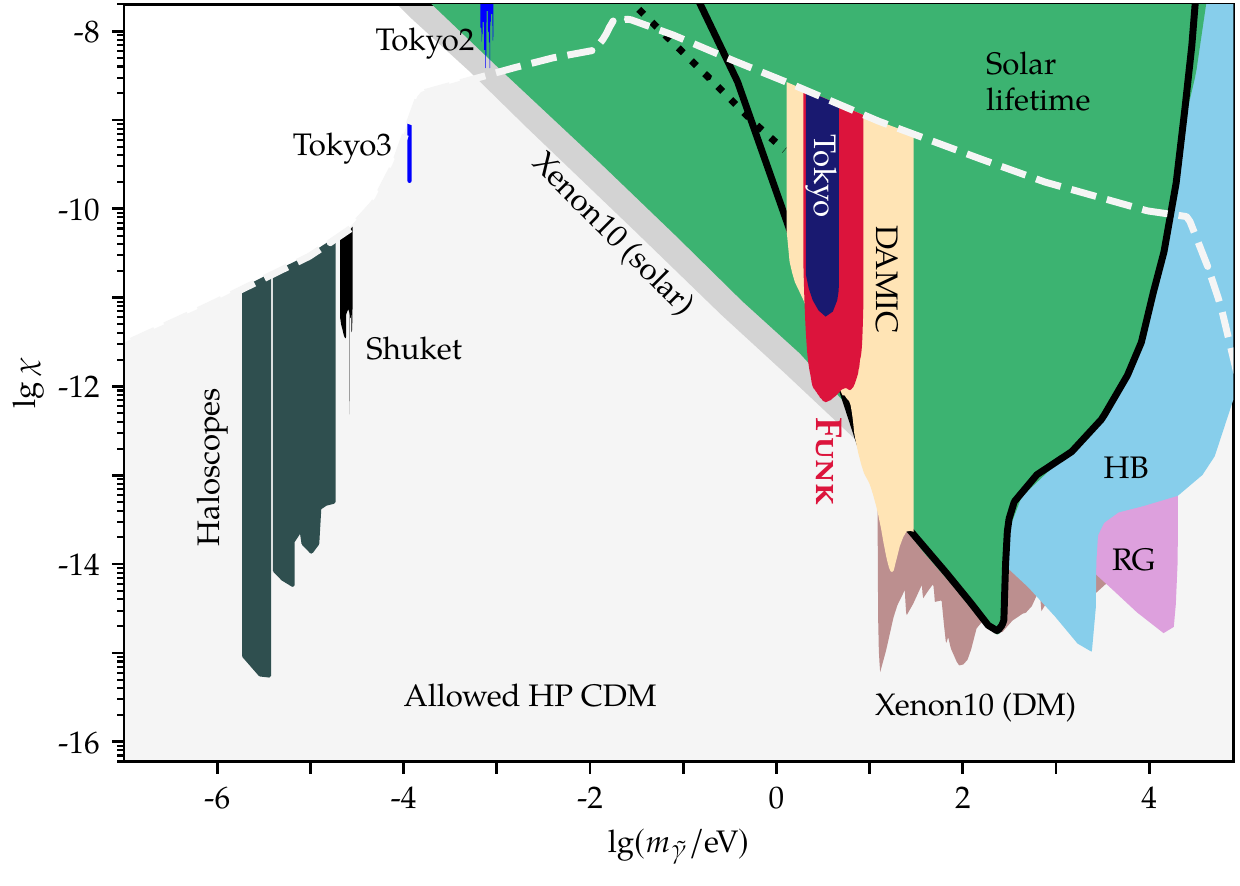}
\caption{\funk exclusion limit at a 95\% C.L.\ (red-shaded region).
The Tokyo limit was obtained by a similar setup but with a smaller mirror~\cite{Suzuki:2015sza}.
HP bounds from other direct-detection experiments are also shown for comparison: DAMIC~\cite{Aguilar-Arevalo:2019wdi}, Xenon10 (both as DM and as dark radiation emitted by the sun)~\cite{Bloch:2016sjj}, Tokyo2~\cite{Knirck:2018ojz},
Tokyo3~\cite{Tomita:2020usq}, Shuket~\cite{Brun:2019kak}, and DM haloscopes~\cite{Arias:2012az}.
The regions labeled Solar lifetime, HB (horizontal-branch stars) and RB (red giants) are indirect constraints derived from astrophysical considerations~\cite{Redondo:2013lna}.
Finally, the grey-shaded region corresponds to the allowed parameter-space for HPs to act as CDM~\cite{Nelson:2011sf,Arias:2012az}.}
\label{f:limit}
\end{figure*}

To account for the physical bound of the rate being positive-definite, we use the Feldman-Cousin prescription to derive a positive upper-limit on the minimum detectable HP signal.
We obtain $r_\text{det}\lesssim0.0238$\,Hz at a 95\%\,C.L.
Inserting this value into \cref{e:signal_value}, we obtain an upper bound on the magnitude of the kinetic-mixing parameter such that $\chi\lesssim6.65{\times}10^{-13}$ at a 95\%\,C.L. for the range of HP masses $1.95<m_{\tilde\gamma}/\text{eV}<8.55$.
The mass-resolved limit is presented in \cref{f:limit_zoom,f:limit}, and is given in tabular form in \cref{s:data}.

\section{Conclusion}

Understanding the nature and origin of DM in the Universe is possibly one of the most urgent tasks in particle physics today.
Measurements with colliders and direct searches for WIMPs have imposed strong constraints on the existence of heavy DM in recent years.
The exploration of the frontier of light and ultralight DM is presently gaining momentum and becoming a crucial direction of research for the near future~\cite{Strategy:2019vxc}.

With \funk, we have set up an experiment that can search for ultralight hidden photons as candidates for dark matter, and we have reported an improved limit in the visible and ultraviolet range of the hidden-photon masses using the dish-antenna method.
In the mass range of 2.5 to 7\,eV, we are able to constrain and exclude HP coupling stronger than $\chi\approx10^{-12}$ in kinetic mixing.
This limit is competitive with those derived from astrophysical considerations (e.g.\ solar lifetime and Xenon10) and, in a certain HP mass range, exceeds those from direct-detection experiments (e.g.\ Tokyo and DAMIC).

One quarter of the background in our measurement is comprised of thermal emission from the PMT cathode, while external sources in the experimental hall and within the field of view of the PMT account for the rest.
The statistical uncertainties are not derived or assumed; instead, we obtain them from direct measurements of the fluctuations in our event rates.
The largest sources of systematic uncertainty arise from reflections on the mirror, a memory effect in the PMT, and a systematic difference in the rates between two modes of measurement.
The magnitudes of each are well determined.
In hindsight, it was extremely useful to include the \emph{closed} shutter modes in our measurement cycle.
These enabled the observation and understanding of the memory effect in our PMT and the development of the correction for this effect, which was applied in our analysis.

Some guidance is available from theory as to what the hidden photon mass might be \cite{Graham:2015rva}, but the mass-range of viable hidden-photon dark matter is huge.
For this reason, techniques which can scan over a large parameter space are especially useful even if they cannot explore couplings as small as is possible with some other methods (e.g.\ resonant haloscopes).
As \funk is using a broadband search technique, a future exploitation of the setup in other wavelength ranges is possible, and the \funk setup therefore continues to carry an enormous physics potential.

\section*{Acknowledgments}

We thank the Helmholtz Alliance for Astroparticle Physics (HAP) for partial funding of the experiment.
A.~A.\ and T.~S.\ acknowledge support from the European Union’s Horizon 2020 research and innovation program under the Marie Sklodowska-Curie grant agreement No.~674896 (Elusives).
A.~A.\ acknowledges the support from the ``Karlsruhe School of Elementary and Astroparticle Physics: Science and Technology'' (KSETA).
B.~D.\ acknowledges support through the European Research Council (ERC) under grant ERC-2018-StG-802836 (AxScale).
We thank the anonymous reviewer for the insightful comments and suggestions.

\appendix

\section{Data}
\label{s:data}

\begin{table}[!h]\centering
\caption{\funk{} exclusion-limit data in tabular form.}
\begin{tabular}{@{}ll@{}}
\toprule
$m_{\tilde{\gamma}}$/eV & $\chi/10^{-13}$
\\
\midrule
1.95 & 245 \\
2.17 & \phantom{0}21.2 \\
2.41 & \phantom{0}10.7 \\
2.68 & \phantom{00}8.71 \\
2.97 & \phantom{00}7.39 \\
3.31 & \phantom{00}6.72 \\
3.67 & \phantom{00}6.91 \\
4.08 & \phantom{00}7.47 \\
4.54 & \phantom{00}8.21 \\
5.04 & \phantom{00}9.24 \\
5.60 & \phantom{00}9.57 \\
6.23 & \phantom{00}9.15 \\
6.92 & \phantom{0}10.7 \\
7.69 & \phantom{0}16.1 \\
8.55 & \phantom{0}74.1 \\
\bottomrule
\end{tabular}
\end{table}

\bibliographystyle{apsrev}
\bibliography{references}

\begin{thebibliography}{59}
\expandafter\ifx\csname natexlab\endcsname\relax\def\natexlab#1{#1}\fi
\expandafter\ifx\csname bibnamefont\endcsname\relax
  \def\bibnamefont#1{#1}\fi
\expandafter\ifx\csname bibfnamefont\endcsname\relax
  \def\bibfnamefont#1{#1}\fi
\expandafter\ifx\csname citenamefont\endcsname\relax
  \def\citenamefont#1{#1}\fi
\expandafter\ifx\csname url\endcsname\relax
  \def\url#1{\texttt{#1}}\fi
\expandafter\ifx\csname urlprefix\endcsname\relax\def\urlprefix{URL }\fi
\providecommand{\bibinfo}[2]{#2}
\providecommand{\eprint}[2][]{\url{#2}}

\bibitem[{\citenamefont{Akrami et~al.}(2018)}]{Akrami:2018vks}
\bibinfo{author}{\bibfnamefont{Y.}~\bibnamefont{Akrami}} \bibnamefont{et~al.}
  (\bibinfo{collaboration}{Planck}) (\bibinfo{year}{2018}),
  \eprint{1807.06205}.

\bibitem[{\citenamefont{Holdom}(1986)}]{Holdom:1985ag}
\bibinfo{author}{\bibfnamefont{B.}~\bibnamefont{Holdom}},
  \bibinfo{journal}{Phys.\ Lett.\ B} \textbf{\bibinfo{volume}{166}},
  \bibinfo{pages}{196} (\bibinfo{year}{1986}).

\bibitem[{\citenamefont{Jaeckel}(2012)}]{Jaeckel:2013ija}
\bibinfo{author}{\bibfnamefont{J.}~\bibnamefont{Jaeckel}},
  \bibinfo{journal}{Frascati Phys.\ Ser.} \textbf{\bibinfo{volume}{56}},
  \bibinfo{pages}{172} (\bibinfo{year}{2012}), \eprint{1303.1821}.

\bibitem[{\citenamefont{Dienes et~al.}(1997)\citenamefont{Dienes, Kolda, and
  March-Russell}}]{Dienes:1996zr}
\bibinfo{author}{\bibfnamefont{K.~R.} \bibnamefont{Dienes}},
  \bibinfo{author}{\bibfnamefont{C.~F.} \bibnamefont{Kolda}}, \bibnamefont{and}
  \bibinfo{author}{\bibfnamefont{J.}~\bibnamefont{March-Russell}},
  \bibinfo{journal}{Nucl.\ Phys.\ B} \textbf{\bibinfo{volume}{492}},
  \bibinfo{pages}{104} (\bibinfo{year}{1997}), \eprint{hep-ph/9610479}.

\bibitem[{Hew(2012)}]{Hewett:2012ns}
\emph{\bibinfo{title}{{Fundamental Physics at the Intensity Frontier}}}
  (\bibinfo{year}{2012}), \eprint{1205.2671},
  \urlprefix\url{http://lss.fnal.gov/archive/preprint/fermilab-conf-12-879-ppd.shtml}.

\bibitem[{\citenamefont{Essig et~al.}(2013)}]{Essig:2013lka}
\bibinfo{author}{\bibfnamefont{R.}~\bibnamefont{Essig}} \bibnamefont{et~al.},
  in \emph{\bibinfo{booktitle}{{Proceedings, 2013 Community Summer Study on the
  Future of U.S. Particle Physics: Snowmass on the Mississippi (CSS2013):
  Minneapolis, MN, USA, July 29-August 6, 2013}}} (\bibinfo{year}{2013}),
  \eprint{1311.0029},
  \urlprefix\url{http://www.slac.stanford.edu/econf/C1307292/docs/IntensityFrontier/NewLight-17.pdf}.

\bibitem[{\citenamefont{Battaglieri et~al.}(2017)}]{Battaglieri:2017aum}
\bibinfo{author}{\bibfnamefont{M.}~\bibnamefont{Battaglieri}}
  \bibnamefont{et~al.}, in \emph{\bibinfo{booktitle}{{U.S. Cosmic Visions: New
  Ideas in Dark Matter College Park, MD, USA, March 23-25, 2017}}}
  (\bibinfo{year}{2017}), \eprint{1707.04591},
  \urlprefix\url{http://lss.fnal.gov/archive/2017/conf/fermilab-conf-17-282-ae-ppd-t.pdf}.

\bibitem[{\citenamefont{Alemany et~al.}(2019)}]{Alemany:2019vsk}
\bibinfo{author}{\bibfnamefont{R.}~\bibnamefont{Alemany}} \bibnamefont{et~al.}
  (\bibinfo{year}{2019}), \eprint{1902.00260}.

\bibitem[{\citenamefont{Beacham et~al.}(2019)}]{Beacham:2019nyx}
\bibinfo{author}{\bibfnamefont{J.}~\bibnamefont{Beacham}} \bibnamefont{et~al.}
  (\bibinfo{year}{2019}), \eprint{1901.09966}.

\bibitem[{\citenamefont{Arkani-Hamed et~al.}(2009)\citenamefont{Arkani-Hamed,
  Finkbeiner, Slatyer, and Weiner}}]{ArkaniHamed:2008qn}
\bibinfo{author}{\bibfnamefont{N.}~\bibnamefont{Arkani-Hamed}},
  \bibinfo{author}{\bibfnamefont{D.~P.} \bibnamefont{Finkbeiner}},
  \bibinfo{author}{\bibfnamefont{T.~R.} \bibnamefont{Slatyer}},
  \bibnamefont{and} \bibinfo{author}{\bibfnamefont{N.}~\bibnamefont{Weiner}},
  \bibinfo{journal}{Phys.\ Rev.\ D} \textbf{\bibinfo{volume}{79}},
  \bibinfo{pages}{015014} (\bibinfo{year}{2009}), \eprint{0810.0713}.

\bibitem[{\citenamefont{Bennett et~al.}(2006)}]{Bennett:2006fi}
\bibinfo{author}{\bibfnamefont{G.~W.} \bibnamefont{Bennett}}
  \bibnamefont{et~al.} (\bibinfo{collaboration}{Muon g-2}),
  \bibinfo{journal}{Phys.\ Rev.\ D} \textbf{\bibinfo{volume}{73}},
  \bibinfo{pages}{072003} (\bibinfo{year}{2006}), \eprint{hep-ex/0602035}.

\bibitem[{\citenamefont{Pospelov}(2009)}]{Pospelov:2008zw}
\bibinfo{author}{\bibfnamefont{M.}~\bibnamefont{Pospelov}},
  \bibinfo{journal}{Phys.\ Rev.\ D} \textbf{\bibinfo{volume}{80}},
  \bibinfo{pages}{095002} (\bibinfo{year}{2009}), \eprint{0811.1030}.

\bibitem[{\citenamefont{Nelson and Scholtz}(2011)}]{Nelson:2011sf}
\bibinfo{author}{\bibfnamefont{A.~E.} \bibnamefont{Nelson}} \bibnamefont{and}
  \bibinfo{author}{\bibfnamefont{J.}~\bibnamefont{Scholtz}},
  \bibinfo{journal}{Phys.\ Rev.\ D} \textbf{\bibinfo{volume}{84}},
  \bibinfo{pages}{103501} (\bibinfo{year}{2011}), \eprint{1105.2812}.

\bibitem[{\citenamefont{Arias et~al.}(2012)\citenamefont{Arias, Cadamuro,
  Goodsell, Jaeckel, Redondo, and Ringwald}}]{Arias:2012az}
\bibinfo{author}{\bibfnamefont{P.}~\bibnamefont{Arias}},
  \bibinfo{author}{\bibfnamefont{D.}~\bibnamefont{Cadamuro}},
  \bibinfo{author}{\bibfnamefont{M.}~\bibnamefont{Goodsell}},
  \bibinfo{author}{\bibfnamefont{J.}~\bibnamefont{Jaeckel}},
  \bibinfo{author}{\bibfnamefont{J.}~\bibnamefont{Redondo}}, \bibnamefont{and}
  \bibinfo{author}{\bibfnamefont{A.}~\bibnamefont{Ringwald}},
  \bibinfo{journal}{JCAP} \textbf{\bibinfo{volume}{1206}}, \bibinfo{pages}{013}
  (\bibinfo{year}{2012}), \eprint{1201.5902}.

\bibitem[{\citenamefont{Graham et~al.}(2016)\citenamefont{Graham, Mardon, and
  Rajendran}}]{Graham:2015rva}
\bibinfo{author}{\bibfnamefont{P.~W.} \bibnamefont{Graham}},
  \bibinfo{author}{\bibfnamefont{J.}~\bibnamefont{Mardon}}, \bibnamefont{and}
  \bibinfo{author}{\bibfnamefont{S.}~\bibnamefont{Rajendran}},
  \bibinfo{journal}{Phys.\ Rev.\ D} \textbf{\bibinfo{volume}{93}},
  \bibinfo{pages}{103520} (\bibinfo{year}{2016}), \eprint{1504.02102}.

\bibitem[{\citenamefont{Agrawal et~al.}(2020)\citenamefont{Agrawal, Kitajima,
  Reece, Sekiguchi, and Takahashi}}]{Agrawal:2018vin}
\bibinfo{author}{\bibfnamefont{P.}~\bibnamefont{Agrawal}},
  \bibinfo{author}{\bibfnamefont{N.}~\bibnamefont{Kitajima}},
  \bibinfo{author}{\bibfnamefont{M.}~\bibnamefont{Reece}},
  \bibinfo{author}{\bibfnamefont{T.}~\bibnamefont{Sekiguchi}},
  \bibnamefont{and}
  \bibinfo{author}{\bibfnamefont{F.}~\bibnamefont{Takahashi}},
  \bibinfo{journal}{Phys.\ Lett.\ B} \textbf{\bibinfo{volume}{801}},
  \bibinfo{pages}{135136} (\bibinfo{year}{2020}), \eprint{1810.07188}.

\bibitem[{\citenamefont{Dror et~al.}(2019)\citenamefont{Dror, Harigaya, and
  Narayan}}]{Dror:2018pdh}
\bibinfo{author}{\bibfnamefont{J.~A.} \bibnamefont{Dror}},
  \bibinfo{author}{\bibfnamefont{K.}~\bibnamefont{Harigaya}}, \bibnamefont{and}
  \bibinfo{author}{\bibfnamefont{V.}~\bibnamefont{Narayan}},
  \bibinfo{journal}{Phys.\ Rev.\ D} \textbf{\bibinfo{volume}{99}},
  \bibinfo{pages}{035036} (\bibinfo{year}{2019}), \eprint{1810.07195}.

\bibitem[{\citenamefont{Co et~al.}(2019)\citenamefont{Co, Pierce, Zhang, and
  Zhao}}]{Co:2018lka}
\bibinfo{author}{\bibfnamefont{R.~T.} \bibnamefont{Co}},
  \bibinfo{author}{\bibfnamefont{A.}~\bibnamefont{Pierce}},
  \bibinfo{author}{\bibfnamefont{Z.}~\bibnamefont{Zhang}}, \bibnamefont{and}
  \bibinfo{author}{\bibfnamefont{Y.}~\bibnamefont{Zhao}},
  \bibinfo{journal}{Phys.\ Rev.\ D} \textbf{\bibinfo{volume}{99}},
  \bibinfo{pages}{075002} (\bibinfo{year}{2019}), \eprint{1810.07196}.

\bibitem[{\citenamefont{Bastero-Gil et~al.}(2019)\citenamefont{Bastero-Gil,
  Santiago, Ubaldi, and Vega-Morales}}]{Bastero-Gil:2018uel}
\bibinfo{author}{\bibfnamefont{M.}~\bibnamefont{Bastero-Gil}},
  \bibinfo{author}{\bibfnamefont{J.}~\bibnamefont{Santiago}},
  \bibinfo{author}{\bibfnamefont{L.}~\bibnamefont{Ubaldi}}, \bibnamefont{and}
  \bibinfo{author}{\bibfnamefont{R.}~\bibnamefont{Vega-Morales}},
  \bibinfo{journal}{JCAP} \textbf{\bibinfo{volume}{1904}}, \bibinfo{pages}{015}
  (\bibinfo{year}{2019}), \eprint{1810.07208}.

\bibitem[{\citenamefont{Long and Wang}(2019)}]{Long:2019lwl}
\bibinfo{author}{\bibfnamefont{A.~J.} \bibnamefont{Long}} \bibnamefont{and}
  \bibinfo{author}{\bibfnamefont{L.-T.} \bibnamefont{Wang}},
  \bibinfo{journal}{Phys.\ Rev.\ D} \textbf{\bibinfo{volume}{99}},
  \bibinfo{pages}{063529} (\bibinfo{year}{2019}), \eprint{1901.03312}.

\bibitem[{\citenamefont{Woollett et~al.}(2015)}]{Woollett:2015gma}
\bibinfo{author}{\bibfnamefont{N.}~\bibnamefont{Woollett}}
  \bibnamefont{et~al.}, in \emph{\bibinfo{booktitle}{{Proceedings, 11th Patras
  Workshop on Axions, WIMPs and WISPs (Axion-WIMP 2015): Zaragoza, Spain, June
  22-26, 2015}}} (\bibinfo{year}{2015}), pp. \bibinfo{pages}{179--182},
  \eprint{1509.07693}.

\bibitem[{\citenamefont{Betz}(2014-01-10)}]{Betz:2014wie}
\bibinfo{author}{\bibfnamefont{M.}~\bibnamefont{Betz}}, Ph.D. thesis,
  \bibinfo{school}{KIT, Karlsruhe} (\bibinfo{year}{2014-01-10}).

\bibitem[{\citenamefont{Ehret et~al.}(2010)}]{Ehret:2010mh}
\bibinfo{author}{\bibfnamefont{K.}~\bibnamefont{Ehret}} \bibnamefont{et~al.},
  \bibinfo{journal}{Phys.\ Lett.\ B} \textbf{\bibinfo{volume}{689}},
  \bibinfo{pages}{149} (\bibinfo{year}{2010}), \eprint{1004.1313}.

\bibitem[{\citenamefont{Redondo}(2015)}]{Redondo:2015iea}
\bibinfo{author}{\bibfnamefont{J.}~\bibnamefont{Redondo}},
  \bibinfo{journal}{JCAP} \textbf{\bibinfo{volume}{1507}}, \bibinfo{pages}{024}
  (\bibinfo{year}{2015}), \eprint{1501.07292}.

\bibitem[{\citenamefont{Schwarz et~al.}(2015)\citenamefont{Schwarz, Knabbe,
  Lindner, Redondo, Ringwald, Schneide, Susol, and
  Wiedemann}}]{Schwarz:2015lqa}
\bibinfo{author}{\bibfnamefont{M.}~\bibnamefont{Schwarz}},
  \bibinfo{author}{\bibfnamefont{E.-A.} \bibnamefont{Knabbe}},
  \bibinfo{author}{\bibfnamefont{A.}~\bibnamefont{Lindner}},
  \bibinfo{author}{\bibfnamefont{J.}~\bibnamefont{Redondo}},
  \bibinfo{author}{\bibfnamefont{A.}~\bibnamefont{Ringwald}},
  \bibinfo{author}{\bibfnamefont{M.}~\bibnamefont{Schneide}},
  \bibinfo{author}{\bibfnamefont{J.}~\bibnamefont{Susol}}, \bibnamefont{and}
  \bibinfo{author}{\bibfnamefont{G.}~\bibnamefont{Wiedemann}},
  \bibinfo{journal}{JCAP} \textbf{\bibinfo{volume}{1508}}, \bibinfo{pages}{011}
  (\bibinfo{year}{2015}), \eprint{1502.04490}.

\bibitem[{\citenamefont{Sikivie}(1983)}]{Sikivie:1983ip}
\bibinfo{author}{\bibfnamefont{P.}~\bibnamefont{Sikivie}},
  \bibinfo{journal}{Phys.\ Rev.\ Lett.} \textbf{\bibinfo{volume}{51}},
  \bibinfo{pages}{1415} (\bibinfo{year}{1983}), \bibinfo{note}{[,321(1983)]}.

\bibitem[{\citenamefont{Hoang~Nguyen et~al.}(2019)\citenamefont{Hoang~Nguyen,
  Lobanov, and Horns}}]{Nguyen:2019xuh}
\bibinfo{author}{\bibfnamefont{L.}~\bibnamefont{Hoang~Nguyen}},
  \bibinfo{author}{\bibfnamefont{A.}~\bibnamefont{Lobanov}}, \bibnamefont{and}
  \bibinfo{author}{\bibfnamefont{D.}~\bibnamefont{Horns}},
  \bibinfo{journal}{JCAP} \textbf{\bibinfo{volume}{1910}}, \bibinfo{pages}{014}
  (\bibinfo{year}{2019}), \eprint{1907.12449}.

\bibitem[{\citenamefont{Silva-Feaver et~al.}(2017)}]{Silva-Feaver:2016qhh}
\bibinfo{author}{\bibfnamefont{M.}~\bibnamefont{Silva-Feaver}}
  \bibnamefont{et~al.}, \bibinfo{journal}{IEEE Trans.\ Appl.\ Supercond.}
  \textbf{\bibinfo{volume}{27}}, \bibinfo{pages}{1400204}
  (\bibinfo{year}{2017}), \eprint{1610.09344}.

\bibitem[{\citenamefont{Arias et~al.}(2015)\citenamefont{Arias, Arza,
  {D\"obrich}, Gamboa, and M\'endez}}]{Arias:2014ela}
\bibinfo{author}{\bibfnamefont{P.}~\bibnamefont{Arias}},
  \bibinfo{author}{\bibfnamefont{A.}~\bibnamefont{Arza}},
  \bibinfo{author}{\bibfnamefont{B.}~\bibnamefont{{D\"obrich}}},
  \bibinfo{author}{\bibfnamefont{J.}~\bibnamefont{Gamboa}}, \bibnamefont{and}
  \bibinfo{author}{\bibfnamefont{F.}~\bibnamefont{M\'endez}},
  \bibinfo{journal}{Eur.\ Phys.\ J.\ C} \textbf{\bibinfo{volume}{75}},
  \bibinfo{pages}{310} (\bibinfo{year}{2015}), \eprint{1411.4986}.

\bibitem[{\citenamefont{Chaudhuri et~al.}(2015)\citenamefont{Chaudhuri, Graham,
  Irwin, Mardon, Rajendran, and Zhao}}]{Chaudhuri:2014dla}
\bibinfo{author}{\bibfnamefont{S.}~\bibnamefont{Chaudhuri}},
  \bibinfo{author}{\bibfnamefont{P.~W.} \bibnamefont{Graham}},
  \bibinfo{author}{\bibfnamefont{K.}~\bibnamefont{Irwin}},
  \bibinfo{author}{\bibfnamefont{J.}~\bibnamefont{Mardon}},
  \bibinfo{author}{\bibfnamefont{S.}~\bibnamefont{Rajendran}},
  \bibnamefont{and} \bibinfo{author}{\bibfnamefont{Y.}~\bibnamefont{Zhao}},
  \bibinfo{journal}{Phys.\ Rev.\ D} \textbf{\bibinfo{volume}{92}},
  \bibinfo{pages}{075012} (\bibinfo{year}{2015}), \eprint{1411.7382}.

\bibitem[{\citenamefont{Veberi\v{c}
  et~al.}(2018{\natexlab{a}})}]{Experiment:2017icw}
\bibinfo{author}{\bibfnamefont{D.}~\bibnamefont{Veberi\v{c}}}
  \bibnamefont{et~al.} (\bibinfo{collaboration}{\textsc{Funk} Experiment}),
  \bibinfo{journal}{PoS} \textbf{\bibinfo{volume}{ICRC2017}},
  \bibinfo{pages}{880} (\bibinfo{year}{2018}{\natexlab{a}}),
  \eprint{1711.02958}.

\bibitem[{\citenamefont{Engel et~al.}(2017)}]{Experiment:2017kmm}
\bibinfo{author}{\bibfnamefont{R.}~\bibnamefont{Engel}} \bibnamefont{et~al.}
  (\bibinfo{collaboration}{\textsc{Funk} Experiment}), in
  \emph{\bibinfo{booktitle}{{Proceedings, 13th Patras Workshop on Axions, WIMPs
  and WISPs, (PATRAS 2017): Thessaloniki, Greece, 15 May 2017 - 19, 2017}}}
  (\bibinfo{year}{2017}), \eprint{1711.02961}.

\bibitem[{\citenamefont{Horns et~al.}(2013)\citenamefont{Horns, Jaeckel,
  Lindner, Lobanov, Redondo, and Ringwald}}]{Horns:2012jf}
\bibinfo{author}{\bibfnamefont{D.}~\bibnamefont{Horns}},
  \bibinfo{author}{\bibfnamefont{J.}~\bibnamefont{Jaeckel}},
  \bibinfo{author}{\bibfnamefont{A.}~\bibnamefont{Lindner}},
  \bibinfo{author}{\bibfnamefont{A.}~\bibnamefont{Lobanov}},
  \bibinfo{author}{\bibfnamefont{J.}~\bibnamefont{Redondo}}, \bibnamefont{and}
  \bibinfo{author}{\bibfnamefont{A.}~\bibnamefont{Ringwald}},
  \bibinfo{journal}{JCAP} \textbf{\bibinfo{volume}{1304}}, \bibinfo{pages}{016}
  (\bibinfo{year}{2013}), \eprint{1212.2970}.

\bibitem[{\citenamefont{Jaeckel and Redondo}(2013)}]{Jaeckel:2013eha}
\bibinfo{author}{\bibfnamefont{J.}~\bibnamefont{Jaeckel}} \bibnamefont{and}
  \bibinfo{author}{\bibfnamefont{J.}~\bibnamefont{Redondo}},
  \bibinfo{journal}{Phys.\ Rev.\ D} \textbf{\bibinfo{volume}{88}},
  \bibinfo{pages}{115002} (\bibinfo{year}{2013}), \eprint{1308.1103}.

\bibitem[{\citenamefont{Jaeckel and Knirck}(2016)}]{Jaeckel:2015kea}
\bibinfo{author}{\bibfnamefont{J.}~\bibnamefont{Jaeckel}} \bibnamefont{and}
  \bibinfo{author}{\bibfnamefont{S.}~\bibnamefont{Knirck}},
  \bibinfo{journal}{JCAP} \textbf{\bibinfo{volume}{1601}}, \bibinfo{pages}{005}
  (\bibinfo{year}{2016}), \eprint{1509.00371}.

\bibitem[{\citenamefont{Veberi\v{c} et~al.}(2016)}]{Veberic:2015yua}
\bibinfo{author}{\bibfnamefont{D.}~\bibnamefont{Veberi\v{c}}}
  \bibnamefont{et~al.} (\bibinfo{collaboration}{\textsc{Funk} Experiment}),
  \bibinfo{journal}{PoS} \textbf{\bibinfo{volume}{ICRC2015}},
  \bibinfo{pages}{1191} (\bibinfo{year}{2016}), \eprint{1509.02386}.

\bibitem[{\citenamefont{D\"obrich et~al.}(2015)}]{Dobrich:2015tpa}
\bibinfo{author}{\bibfnamefont{B.}~\bibnamefont{D\"obrich}}
  \bibnamefont{et~al.}, in \emph{\bibinfo{booktitle}{{Photon 2015:
  International Conference on the Structure and Interactions of the Photon and
  21th International Workshop on Photon-Photon Collisions and International
  Workshop on High Energy Photon Linear Colliders Novosibirsk, Russia, June
  15-19, 2015}}} (\bibinfo{year}{2015}), \eprint{1510.05869}.

\bibitem[{\citenamefont{Suzuki et~al.}(2015)\citenamefont{Suzuki, Horie, Inoue,
  and Minowa}}]{Suzuki:2015sza}
\bibinfo{author}{\bibfnamefont{J.}~\bibnamefont{Suzuki}},
  \bibinfo{author}{\bibfnamefont{T.}~\bibnamefont{Horie}},
  \bibinfo{author}{\bibfnamefont{Y.}~\bibnamefont{Inoue}}, \bibnamefont{and}
  \bibinfo{author}{\bibfnamefont{M.}~\bibnamefont{Minowa}},
  \bibinfo{journal}{JCAP} \textbf{\bibinfo{volume}{1509}}, \bibinfo{pages}{042}
  (\bibinfo{year}{2015}), \eprint{1504.00118}.

\bibitem[{\citenamefont{Knirck et~al.}(2018)\citenamefont{Knirck, Yamazaki,
  Okesaku, Asai, Idehara, and Inada}}]{Knirck:2018ojz}
\bibinfo{author}{\bibfnamefont{S.}~\bibnamefont{Knirck}},
  \bibinfo{author}{\bibfnamefont{T.}~\bibnamefont{Yamazaki}},
  \bibinfo{author}{\bibfnamefont{Y.}~\bibnamefont{Okesaku}},
  \bibinfo{author}{\bibfnamefont{S.}~\bibnamefont{Asai}},
  \bibinfo{author}{\bibfnamefont{T.}~\bibnamefont{Idehara}}, \bibnamefont{and}
  \bibinfo{author}{\bibfnamefont{T.}~\bibnamefont{Inada}},
  \bibinfo{journal}{JCAP} \textbf{\bibinfo{volume}{1811}}, \bibinfo{pages}{031}
  (\bibinfo{year}{2018}), \eprint{1806.05120}.

\bibitem[{\citenamefont{Brun et~al.}(2019)\citenamefont{Brun, Chevalier, and
  Flouzat}}]{Brun:2019kak}
\bibinfo{author}{\bibfnamefont{P.}~\bibnamefont{Brun}},
  \bibinfo{author}{\bibfnamefont{L.}~\bibnamefont{Chevalier}},
  \bibnamefont{and} \bibinfo{author}{\bibfnamefont{C.}~\bibnamefont{Flouzat}},
  \bibinfo{journal}{Phys.\ Rev.\ Lett.} \textbf{\bibinfo{volume}{122}},
  \bibinfo{pages}{201801} (\bibinfo{year}{2019}), \eprint{1905.05579}.

\bibitem[{\citenamefont{Abraham et~al.}(2004)}]{Abraham:2004dt}
\bibinfo{author}{\bibfnamefont{J.}~\bibnamefont{Abraham}} \bibnamefont{et~al.}
  (\bibinfo{collaboration}{Pierre Auger}), \bibinfo{journal}{Nucl.\ Instrum.\
  Meth.\ A} \textbf{\bibinfo{volume}{523}}, \bibinfo{pages}{50}
  (\bibinfo{year}{2004}).

\bibitem[{\citenamefont{Aab et~al.}(2015)}]{ThePierreAuger:2015rma}
\bibinfo{author}{\bibfnamefont{A.}~\bibnamefont{Aab}} \bibnamefont{et~al.}
  (\bibinfo{collaboration}{Pierre Auger}), \bibinfo{journal}{Nucl.\ Instrum.\
  Meth.\ A} \textbf{\bibinfo{volume}{798}}, \bibinfo{pages}{172}
  (\bibinfo{year}{2015}), \eprint{1502.01323}.

\bibitem[{\citenamefont{Abraham et~al.}(2010)}]{Abraham:2009pm}
\bibinfo{author}{\bibfnamefont{J.}~\bibnamefont{Abraham}} \bibnamefont{et~al.}
  (\bibinfo{collaboration}{Pierre Auger}), \bibinfo{journal}{Nucl.\ Instrum.\
  Meth.\ A} \textbf{\bibinfo{volume}{620}}, \bibinfo{pages}{227}
  (\bibinfo{year}{2010}), \eprint{0907.4282}.

\bibitem[{\citenamefont{{ET Enterprises}}({\natexlab{a}})}]{et-manual}
\bibinfo{author}{\bibnamefont{{ET Enterprises}}}, \emph{\bibinfo{title}{{9107B
  series data sheet}}},
  \urlprefix\url{http://et-enterprises.com/images/data_sheets/9107B.pdf}.

\bibitem[{\citenamefont{D\"obrich et~al.}(2014)\citenamefont{D\"obrich,
  Daumiller, Engel, Kowalski, Lindner, Redondo, and Roth}}]{Dobrich:2014kda}
\bibinfo{author}{\bibfnamefont{B.}~\bibnamefont{D\"obrich}},
  \bibinfo{author}{\bibfnamefont{K.}~\bibnamefont{Daumiller}},
  \bibinfo{author}{\bibfnamefont{R.}~\bibnamefont{Engel}},
  \bibinfo{author}{\bibfnamefont{M.}~\bibnamefont{Kowalski}},
  \bibinfo{author}{\bibfnamefont{A.}~\bibnamefont{Lindner}},
  \bibinfo{author}{\bibfnamefont{J.}~\bibnamefont{Redondo}}, \bibnamefont{and}
  \bibinfo{author}{\bibfnamefont{M.}~\bibnamefont{Roth}}, in
  \emph{\bibinfo{booktitle}{{Proceedings, 10th Patras Workshop on Axions, WIMPs
  and WISPs (AXION-WIMP 2014): Geneva, Switzerland, June 29-July 4, 2014}}}
  (\bibinfo{year}{2014}), pp. \bibinfo{pages}{173--176}, \eprint{1410.0200},
  \urlprefix\url{http://motherboard.vice.com/read/a-giant-mirror-to-capture-dark-matter-hidden-light}.

\bibitem[{\citenamefont{{ET Enterprises}}({\natexlab{b}})}]{et-fact50}
\bibinfo{author}{\bibnamefont{{ET Enterprises}}},
  \emph{\bibinfo{title}{{FACT50/LCT50 data sheet}}},
  \urlprefix\url{http://et-enterprises.com/images/data_sheets/FACT50_LCT50.pdf}.

\bibitem[{\citenamefont{Veberi\v{c}
  et~al.}(2018{\natexlab{b}})}]{Veberic:2017icw}
\bibinfo{author}{\bibfnamefont{D.}~\bibnamefont{Veberi\v{c}}}
  \bibnamefont{et~al.} (\bibinfo{collaboration}{\textsc{Funk} Experiment}),
  \bibinfo{journal}{PoS} \textbf{\bibinfo{volume}{ICRC2017}},
  \bibinfo{pages}{880} (\bibinfo{year}{2018}{\natexlab{b}}),
  \eprint{1711.02958}.

\bibitem[{\citenamefont{Coates}(1972)}]{Coates:1972}
\bibinfo{author}{\bibfnamefont{P.~B.} \bibnamefont{Coates}},
  \bibinfo{journal}{J.\ Phys.\ D Appl.\ Phys.} \textbf{\bibinfo{volume}{5}},
  \bibinfo{pages}{1489} (\bibinfo{year}{1972}).

\bibitem[{\citenamefont{Badino}(2000)}]{Badino:2000}
\bibinfo{author}{\bibfnamefont{G.}~\bibnamefont{Badino}},
  \bibinfo{journal}{Int.\ J.\ Speleol.\ B} \textbf{\bibinfo{volume}{29}},
  \bibinfo{pages}{89} (\bibinfo{year}{2000}).

\bibitem[{\citenamefont{Helbing et~al.}(2003)\citenamefont{Helbing,
  Goldschmidt, K{\"o}pke, Matis, Nygren, Przybylski, and
  Stokstad}}]{helbing2003light}
\bibinfo{author}{\bibfnamefont{K.}~\bibnamefont{Helbing}},
  \bibinfo{author}{\bibfnamefont{A.}~\bibnamefont{Goldschmidt}},
  \bibinfo{author}{\bibfnamefont{L.}~\bibnamefont{K{\"o}pke}},
  \bibinfo{author}{\bibfnamefont{H.~S.} \bibnamefont{Matis}},
  \bibinfo{author}{\bibfnamefont{D.~R.} \bibnamefont{Nygren}},
  \bibinfo{author}{\bibfnamefont{G.~T.} \bibnamefont{Przybylski}},
  \bibnamefont{and} \bibinfo{author}{\bibfnamefont{R.~G.}
  \bibnamefont{Stokstad}} (\bibinfo{collaboration}{AMANDA}),
  \bibinfo{type}{Tech. Rep.}, \bibinfo{institution}{AMANDA-IR/20030701}
  (\bibinfo{year}{2003}).

\bibitem[{\citenamefont{Aartsen et~al.}(2017)}]{Aartsen:2016nxy}
\bibinfo{author}{\bibfnamefont{M.~G.} \bibnamefont{Aartsen}}
  \bibnamefont{et~al.} (\bibinfo{collaboration}{IceCube}),
  \bibinfo{journal}{JINST} \textbf{\bibinfo{volume}{12}},
  \bibinfo{pages}{P03012} (\bibinfo{year}{2017}), \eprint{1612.05093}.

\bibitem[{\citenamefont{Molenberghs et~al.}(2010)\citenamefont{Molenberghs,
  Verbeke, Dem\'etrio, and Vieira}}]{Molenberghs:2010}
\bibinfo{author}{\bibfnamefont{G.}~\bibnamefont{Molenberghs}},
  \bibinfo{author}{\bibfnamefont{G.}~\bibnamefont{Verbeke}},
  \bibinfo{author}{\bibfnamefont{C.~G.~B.} \bibnamefont{Dem\'etrio}},
  \bibnamefont{and} \bibinfo{author}{\bibfnamefont{A.~M.~C.}
  \bibnamefont{Vieira}}, \bibinfo{journal}{Stat.\ Sci.}
  \textbf{\bibinfo{volume}{25}}, \bibinfo{pages}{325} (\bibinfo{year}{2010}),
  ISSN \bibinfo{issn}{08834237}.

\bibitem[{\citenamefont{Boziev}(1993)}]{boziev1993}
\bibinfo{author}{\bibfnamefont{S.}~\bibnamefont{Boziev}}, in
  \emph{\bibinfo{booktitle}{Proceedings of 23rd International Cosmic Ray
  Conference, Calgary, Canada, July 19-30, 1993}} (\bibinfo{year}{1993}),
  vol.~\bibinfo{volume}{4}, pp. \bibinfo{pages}{403--406}.

\bibitem[{\citenamefont{Andrianavalomahefa}(2020)}]{arnaud_phd}
\bibinfo{author}{\bibfnamefont{A.}~\bibnamefont{Andrianavalomahefa}}, Ph.D.
  thesis, \bibinfo{school}{Fakult\"at f\"ur Physik, Karlsruher Institut f\"ur
  Technologie (KIT)} (\bibinfo{year}{2020}).

\bibitem[{\citenamefont{Aguilar-Arevalo
  et~al.}(2019)}]{Aguilar-Arevalo:2019wdi}
\bibinfo{author}{\bibfnamefont{A.}~\bibnamefont{Aguilar-Arevalo}}
  \bibnamefont{et~al.} (\bibinfo{collaboration}{DAMIC}) (\bibinfo{year}{2019}),
  \eprint{1907.12628}.

\bibitem[{\citenamefont{Redondo and Raffelt}(2013)}]{Redondo:2013lna}
\bibinfo{author}{\bibfnamefont{J.}~\bibnamefont{Redondo}} \bibnamefont{and}
  \bibinfo{author}{\bibfnamefont{G.}~\bibnamefont{Raffelt}},
  \bibinfo{journal}{JCAP} \textbf{\bibinfo{volume}{1308}}, \bibinfo{pages}{034}
  (\bibinfo{year}{2013}), \eprint{1305.2920}.

\bibitem[{\citenamefont{Bloch et~al.}(2017)\citenamefont{Bloch, Essig, Tobioka,
  Volansky, and Yu}}]{Bloch:2016sjj}
\bibinfo{author}{\bibfnamefont{I.~M.} \bibnamefont{Bloch}},
  \bibinfo{author}{\bibfnamefont{R.}~\bibnamefont{Essig}},
  \bibinfo{author}{\bibfnamefont{K.}~\bibnamefont{Tobioka}},
  \bibinfo{author}{\bibfnamefont{T.}~\bibnamefont{Volansky}}, \bibnamefont{and}
  \bibinfo{author}{\bibfnamefont{T.-T.} \bibnamefont{Yu}},
  \bibinfo{journal}{JHEP} \textbf{\bibinfo{volume}{06}}, \bibinfo{pages}{087}
  (\bibinfo{year}{2017}), \eprint{1608.02123}.

\bibitem[{\citenamefont{Tomita et~al.}(2020)\citenamefont{Tomita, Oguri, Inoue,
  Minowa, Nagasaki, Suzuki, and Tajima}}]{Tomita:2020usq}
\bibinfo{author}{\bibfnamefont{N.}~\bibnamefont{Tomita}},
  \bibinfo{author}{\bibfnamefont{S.}~\bibnamefont{Oguri}},
  \bibinfo{author}{\bibfnamefont{Y.}~\bibnamefont{Inoue}},
  \bibinfo{author}{\bibfnamefont{M.}~\bibnamefont{Minowa}},
  \bibinfo{author}{\bibfnamefont{T.}~\bibnamefont{Nagasaki}},
  \bibinfo{author}{\bibfnamefont{J.}~\bibnamefont{Suzuki}}, \bibnamefont{and}
  \bibinfo{author}{\bibfnamefont{O.}~\bibnamefont{Tajima}}
  (\bibinfo{year}{2020}), \eprint{2006.02828}.

\bibitem[{\citenamefont{Ellis et~al.}(2019)}]{Strategy:2019vxc}
\bibinfo{author}{\bibfnamefont{R.~K.} \bibnamefont{Ellis}} \bibnamefont{et~al.}
  (\bibinfo{year}{2019}), \eprint{1910.11775}.

\end{thebibliography}

\end{document}